\documentclass[11pt]{article}

\pdfoutput=1

\usepackage{hyperref}
\usepackage{amsmath}
\usepackage{amssymb}
\usepackage{epsfig}
\usepackage{cite}
\usepackage{color,colordvi}

\newcommand{\eq}{\begin{equation}}
\newcommand{\eqx}{\end{equation}}
\newcommand{\eqn}{\begin{eqnarray}}
\newcommand{\bi}{\begin{itemize}}
\newcommand{\eqnx}{\end{eqnarray}}
\newcommand{\ei}{\end{itemize}}
\newcommand{\lsim}{\stackrel{<}{_\sim}}

\newcounter{hran}

\newcommand{\be}{\begin{equation}}
\newcommand{\ee}{\end{equation}}
\newcommand{\no}{\nonumber}
\newcommand{\bea}{\begin{eqnarray}}
\newcommand{\eea}{\end{eqnarray}}
\newcommand{\SM}{{\rm SM}}
\newcommand{\NP}{{\rm NP}}
\newcommand{\ntree}{{\rm peng}}

\def\simlt{\stackrel{<}{{}_\sim}}

\begin{document}
\begin{titlepage}
\begin{center}

\hfill CERN-PH-TH/2012-017
\vskip 0.4 cm

\vspace{1cm}

{\Large\bf Direct CP violation in charm and flavor mixing beyond the SM}

\vspace{0.2cm}

\end{center}

\begin{center}

{\bf Gian Francesco Giudice}$^{a}$, {\bf Gino Isidori}$^{a,b}$,  {\bf Paride Paradisi}$^{a}$ 

\vspace{0.4cm}

{\em $^a$CERN,
Theory Division,
1211 Geneva 23, Switzerland}


{\em $^b$INFN, 
Laboratori Nazionali di Frascati, Via E. Fermi 40, 00044 Frascati, Italy}

\end{center}

\vskip .2in

\setlength{\baselineskip}{0.2in}

\centerline{\bf Abstract}
\vskip .1in

\noindent


We analyze possible interpretations of the recent LHCb evidence for CP violation in $D$ meson decays in terms of physics beyond the Standard Model.
On general grounds, models in which the primary source of flavor 
violation is linked to the breaking of chiral symmetry 
(left-right flavor mixing) are natural candidates to explain this effect,
via enhanced chromomagnetic operators.  
In the case of supersymmetric models, we identify two motivated scenarios: 
disoriented $A$-terms and split families. These structures predict other non-standard
signals, such as nuclear EDMs close to their present bounds and,  possibly, 
tiny but visible deviations in $K$ and $B$ physics, or even sizable flavor-violating processes involving the top quark or the stops. Some of these connections, especially 
the one with nuclear EDMs, hold beyond supersymmetry,
as illustrated with the help of prototype non-supersymmetric  models.


\end{titlepage}

\vskip .4in
\noindent

\section{Introduction}

Processes involving $K$ and $B$ mesons have always been regarded as the most interesting probe of flavor and CP violation. Indeed, within the Standard Model (SM), the largest flavor and CP violating effects reside in systems involving down-type quarks, since the top mass is the main source of flavor violation and charged-current loops are needed to communicate symmetry breaking, in agreement with the GIM mechanism. Similarly, sizable CP violations in the SM are always accompanied by flavor transitions.

While these properties hold in the SM, there is no good reason for them to be true if new physics is present at the electroweak scale. In particular, it is quite plausible that new-physics contributions affect mostly the up-type sector, possibly in association with the mechanism responsible for the large top mass. Supersymmetric models with squark alignment~\cite{Nir:1993mx,Leurer:1993gy} provide one example of theories with large flavor and CP violation in the up sector but, as emphasized also in ref.~\cite{Giudice:2011ak}, this situation is fairly general in classes of models in which the flavor hierarchies are explained without invoking the hypothesis of minimal flavor violation~\cite{MFV}. So $D$-meson decays represent a unique probe of new-physics flavor effects, quite complementary to tests in $K$ and $B$ systems.

The LHCb collaboration has recently announced a first evidence for CP violation in charm. 
The difference of the time-integrated CP asymmetries in the decays
$D^0 \to K^+K^-$ and $D^0 \to \pi^+\pi^-$ has been measured to be~\cite{lhcb} 
\begin{equation}
 a_{K^+ K^-} - a_{\pi^+ \pi^-}  = -(0.82\pm 0.21\pm0.11)\%\, ,
\label{LHCbnews}
\end{equation}
where
\begin{equation}
a_f \equiv \frac{\Gamma(D^0\to f)-\Gamma(\bar D^0\to f)}
  {\Gamma(D^0\to f)+\Gamma(\bar D^0\to f)}\, , ~~f=K^+ K^-,\pi^+ \pi^-.
\label{eq:af_def}
\end{equation}
Combining the result in eq.~(\ref{LHCbnews}) with previous measurements of these CP
asymmetries~\cite{Aaltonen:2011se, Staric:2008rx, Aubert:2007if, HFAG} and taking into 
account the contamination of indirect CP violation due to 
the different proper-time cuts in the two decay modes (see below), one finds
a $3.6\sigma$ evidence for a non-vanishing direct CP violating (DCPV) asymmetry:
\begin{equation}
\Delta a_{CP} = a^{\rm dir}_K - a^{\rm dir}_\pi = -(0.65\pm 0.18)\%\,.
\label{eq:acpExp}
\end{equation}

Unfortunately, precise theoretical predictions in $D$-meson decays are notoriously difficult because the charm quark is too heavy for chiral perturbation to be applicable and too light for heavy-quark effective theory to be trusted. Nevertheless, according to the most 
accurate estimates performed before the LHCb measurement~\cite{Grossman:2006jg},  the result in eq.~(\ref{eq:acpExp})  
is larger than the SM expectation and requires an enhancement of the CKM-suppressed amplitudes by about a factor of 
5--10~\cite{Isidori:2011qw}. At the moment it is impossible to argue that such an enhancement is not present already in the SM, as claimed 
long ago in ref.~\cite{Golden:1989qx} and stressed recently in ref.~\cite{Brod:2011re}.  
However, it is natural to start speculating about the implications of this measurement, under the hypothesis that new physics is responsible for (at least part of)  
the effect. A first important step in this direction has been made in ref.~\cite{Isidori:2011qw}, where the new-physics interpretation was expressed in terms of effective operators. Our aim in this paper is to pursue the analysis, searching for specific theories at the electroweak scale that give rise to the effective operators identified in ref.~\cite{Isidori:2011qw}.

Using the results of~\cite{Isidori:2011qw}, we can argue that a large new-physics (imaginary) contribution to the $\Delta C=1$ chromomagnetic operator is the best candidate to explain the LHCb result, while being compatible with all current data in flavor physics. Indeed, a $\Delta C=1$ four-fermion operator is not very promising because, inserted twice in a quadratically-divergent loop of light quarks, it generates a $\Delta C=2$ four-fermion interaction, typically leading to an exceedingly large contribution to $D^0-\bar D^0$ mixing. Moreover a $\Delta C=1$ four-fermion operator involving at least two left-handed quarks, when dressed by $W$ exchange, generates a $\Delta S=1$ interaction potentially dangerous for $\epsilon^\prime /\epsilon$. On the other hand, the $\Delta C=1$ chromomagnetic operator has a coefficient suppressed at least by a charm Yukawa coupling, because of the necessary chiral transition. In $D$-meson decays, this Yukawa suppression is exactly compensated by an enhancement in the matrix element of a factor $v/m_c$. Here the Higgs vacuum expectation value $v$ comes from the structure of the gauge-invariant dimension-six operator and we are not distinguishing between meson and quark masses ($m_D \approx m_c$). When the $\Delta C=1$ chromomagnetic operator is dressed to induce a $\Delta C=2$ operator, the Yukawa suppression cannot be compensated by the four-fermion matrix element and thus the effect on $D^0-\bar D^0$ mixing is always suppressed at least by a factor $m_c^2/v^2$. Similarly, $\Delta S =1$ transitions cannot be generated by virtual $W$ exchange without a light quark mass insertion, which is always needed because of the left-right structure of the chromomagnetic interaction. As a result, contributions to $\epsilon^\prime /\epsilon$ are suppressed at least by the square of the charm Yukawa coupling.

The challenge of model building is to generate the $\Delta C=1$ chromomagnetic operator without inducing dangerous 4-fermion operators that lead to unacceptably large effects in $D^0-\bar D^0$ mixing or in flavor processes in the down-type quark sector. In 
sections~\ref{sect:SUSY} and \ref{sect:nonSUSY} we show that various classes of models naturally satisfy these conditions and can properly explain the LHCb observation, both in the context of supersymmetric theories and of theories with $Z$ or scalar flavor-violating interactions. 

We also point out connections between the CP violation in charm reported by LHCb and other independent observables. New measurements of these observables can provide us with hints in favor or against a new-physics interpretation of the LHCb result and can help us to discriminate among different new-physics models. Especially interesting are the 
electric dipole moments (EDMs) of the neutron and the mercury, which happen to be very close to their present bounds in all the framework considered. In most models  flavor-changing top-quark decays are strongly enhanced over the SM predictions, although not always at a detectable level. More model-dependent connections include possible small deviations from the SM in $B_d$ meson mixing and/or in rare $B$ and $K$ decays, as well as flavor-violating effects in squark production and decays (in the case of supersymmetry).

The paper is organized as follows. In sect.~\ref{sec:charm} we introduce the main formulae to evaluate 
$\Delta a_{CP} $ and $D^0$--$\bar D^0$ mixing, both within and beyond the SM. Similarly, in sect.~\ref{sec:Other} 
we introduce the general formalism relevant to $B$--$\bar B$ mixing and rare top decays. 
Section~\ref{sect:SUSY} is devoted to supersymmetry, where we distinguish three main favor structures: disoriented $A$ terms, 
alignment, and split families. In sect.~\ref{sect:nonSUSY} we analyze the case of theories with $Z$ or Higgs 
flavor-violating interactions. The results are summarized in the conclusions.

\section{CP violation in the charm system}
\label{sec:charm}

\subsection{Direct CP violation in $D\to\pi\pi,KK$}
\label{sec:DCPC}

The singly-Cabibbo-suppressed decay amplitude $A_f\ (\bar A_f)$ of $D^0\ (\bar D^0)$ to a CP eigenstate
$f$ can be decomposed as~\cite{Grossman:2006jg}
\begin{subequations}
\begin{eqnarray}
A_f &=& A^T_f\, e^{i \phi_f^T} \big[1+r_f\, e^{i(\delta_f + \phi_f)}\big]\,, \\
\bar A_f &=& \eta_{CP}\, A^T_f\, e^{-i \phi_f^T}
  \big[1+r_f\, e^{i(\delta_f - \phi_f)}\big]\,,
\end{eqnarray}
\end{subequations}
where $ \eta_{CP}=\pm1$ is the CP eigenvalue of the final-state $f$.
Magnitude and weak phase of the the dominant amplitude
are denoted by $A^T$ and $\phi^T_f$, while 
$r_f$ parameterizes the relative magnitude of all the subleading amplitudes
with different strong ($\delta_f$) and weak ($\phi_f$) phases relative 
to the leading term. A necessary condition for a non-vanishing 
DCPV asymmetry is that $r_f$, $\delta$ and $\phi_f$ are all different than zero.
Indeed, in the limit where $r_f\ll 1$, which is an excellent approximation given the 
experimental size of DCPV, 
\be
a_f^{\rm dir} \equiv \frac{\left| A_f \right|^2 -\left| \bar A_f \right|^2  }{  \left| A_f \right|^2 + \left| \bar A_f \right|^2  } 
 =  -2 {r_f} \sin \delta_f \sin \phi_f~.
 \label{eq:Adirdef}
\ee

The $\Delta C=1$ effective weak Hamiltonian describing $D$-meson decays
within the SM, renormalized at a low scale ($m_c < \mu< m_b$), 
can be decomposed as
\be
\label{Heff}
\mathcal H^{\rm eff}_{\Delta C = 1} =  \sum_{q=d,s} 
  \lambda_q\, \mathcal H^{q}_{\Delta C = 1}
  + \lambda_b\,  \mathcal H^{\ntree}_{\Delta C = 1} + {\rm h.c.},
\ee
where $\lambda_i = V_{ci}^* V_{ui}$  are the relevant 
CKM factors and  $\mathcal H^{i}_{|\Delta c| = 1}$ denote a series 
of dimension-six operators written in terms of light SM fields 
(see~e.g.~\cite{Grossman:2006jg} for more details).
Making use of the CKM unitarity  relation $\lambda_d + \lambda_s + \lambda_b=0$, 
one can write $A_{K} = \lambda_s (A^s_{K} - A^d_{K}) + \lambda_b
(A_{K}^b - A_{K}^d)$ and $A_\pi = \lambda_d (A^d_\pi - A^s_\pi) + \lambda_b
(A_\pi^b - A_\pi^s)$ such that the first terms are singly-Cabibbo-suppressed,
while the second terms have a much stronger CKM suppression 
and have either vanishing tree-level matrix elements or tiny Wilson coefficients.
The magnitudes of these subleading amplitudes are controlled by the CKM
ratio $|\lambda_b/\lambda_{s,d} | \approx 7\times 10^{-4}$
and by the following ratios of hadronic amplitudes~\cite{Isidori:2011qw}:
\be\label{RKpidef}
R^{\SM}_K = \frac{ A^b_K - A^d_K }{ A_K^s - A_K^d }\,, \qquad
R^{\SM}_\pi = \frac{ A^b_\pi - A^s_\pi }{ A_\pi^d - A_\pi^s }\,.
\ee

Similarly, new-physics effects can be described in full generality by an effective 
Hamiltonian of the type 
\be
\mathcal H^{\rm eff-\rm NP}_{|\Delta c| = 1} = \frac{G_F}{\sqrt 2} 
  \sum_{i } C_i  Q_i + {\rm h.c.}\,,
\label{eq:HNP}
\ee
where the $Q_i$ are dimension-six effective operators written in terms of light SM fields 
(see~e.g.~\cite{Isidori:2011qw} for the complete list) renormalized at the low scale and $C_i$ are the corresponding Wilson coefficients.
As anticipated in the introduction, we are particularly interested in  
\bea
Q_{8} &=&  \frac{m_c}{4\pi^2}\, \bar u_L \sigma_{\mu\nu} 
    T^a g_s G_a^{\mu\nu} c_R \,, \nonumber  \\
\tilde Q_{8} &=& \frac{m_c}{4\pi^2}\, \bar u_R \sigma_{\mu\nu} 
   T^a g_s G_a^{\mu\nu} c_L \,.
   \label{eq:Q8def}
\eea
The charm Yukawa factor in the normalization of $\tilde Q_{8}$ is only a convention chosen for later convenience, since the natural chiral factor would be the up-quark Yukawa. 

The new-physics amplitudes are then decomposed as 
$A^{\rm NP}_f = \sum C_i A^i_f$ and, in analogy to~eq.~(\ref{RKpidef}), 
we define
\be
R^{{\NP}_i}_K = \frac{ A^i_K}{ A_K^s - A_K^d }\,, \qquad
R^{{\NP}_i}_\pi = \frac{ A^i_\pi }{ A_\pi^d - A_\pi^s }\,.
\label{eq:RpiKNP}
\ee
With these definitions, we obtain~\cite{Isidori:2011qw} 
\bea
\Delta a_{CP} &\approx& \frac{-2}{\sin \theta_c} \left[ \mathrm{Im} (V_{cb}^*V_{ub}) \mathrm{Im} (\Delta R^{\SM}) + \sum_i \mathrm{Im} (C_i^{\NP})\, \mathrm{Im} (\Delta R^{{\NP}_{i}} )\right]
\nonumber \\
&=& -(0.13\%) \mathrm{Im} (\Delta R^{\SM})  
  -9 \sum_i \mathrm{Im} (C_i^{\NP})\, \mathrm{Im} (\Delta R^{{\NP}_{i}} )\,,
\label{eq:acpNP}
\eea
where $\sin \theta_c$ is the Cabibbo angle and $\Delta R^{\SM,{\NP_i}} = R_K^{\SM,{\NP_i}}  + R^{\SM,{\NP_i}} _\pi$. Equation~(\ref{eq:acpNP}) shows that the SM can account for the result in eq.~(\ref{eq:acpExp}) only if $\mathrm{Im} (\Delta R^{\SM})\approx 5$. A naive estimate in perturbation theory gives $\Delta R^{\SM} \approx \alpha_s (m_c)/\pi \approx 0.1$, but a much larger result from non-perturbative effects is in general expected. 

In the $SU(3)$ limit $R_K^{\SM} = R_\pi^{\SM}$, 
hence within the SM $a_K^{\rm dir}$ and $a_\pi^{\rm dir}$ 
should add constructively in $\Delta a_{CP}$~\cite{Golden:1989qx, Quigg:1979ic}.
However, we recall that the observed decay rates 
of $D^0 \to K^+K^-, \pi^+\pi^-$ exhibit  $SU(3)$ breaking effects 
around the $30$--$40\%$ level. If the leading new-physics 
contributions are generated by the chromomagnetic 
operators in eq.~(\ref{eq:Q8def}), it remains true that 
$a_K^{\rm dir}  = - a_\pi^{\rm dir}$ in the  $SU(3)$ limit. 
Hereafter we assume $a_K^{\rm dir}  = - a_\pi^{\rm dir}$. 

The values of $\Delta R^{\NP_i} $ for the 
two chromomagnetic operators in eq.~(\ref{eq:Q8def}) can be estimated 
using naive factorization as in ref.~\cite{Grossman:2006jg}. We find
\be
 R_{K,\pi}^{8, \tilde 8} =    \frac{2 \alpha_s }{ 9 \pi a_1}  
\left( r_\chi + I_\phi \right)  \approx 0.1~, 
\label{eq:a_CP_th}
\ee
where $a_1=  C_1 + C_2/N_c \approx 1$ and, following  ref.~\cite{Grossman:2006jg}, 
we have set $\alpha_s/\pi \approx 0.1$, $r_\chi = 2 m^2_K/(m_s m_c) \approx  2.5$
and $I_\phi = 3$ for the integral of the leading-twist light-cone distribution
amplitude of the $D$ meson. Assuming maximal strong phases, this implies
\be
\left| \mathrm{Im} (\Delta R^{{\NP}_{8, \tilde 8}}) \right| \approx 0.2~.
\label{blabla}
\ee
In the following we use this value as reference estimate
for our numerical analyses,  keeping in mind that it is affected by O(1) uncertainties.

\subsection{CP violation in $D^0 -\bar D^0$ mixing}
\label{sect:DDmix}

The $D^0$--$\bar D^0$ transition amplitude can be decomposed into a 
dispersive ($M_{12}$) and an absorptive ($\Gamma_{12}$) component: 
\begin{eqnarray}
\langle D^0 |\mathcal{H}_{\rm eff}| \bar D^0 \rangle &=& M^D_{12} - \frac{i}{2}\Gamma^D_{12} ~.
\end{eqnarray}
The weak phases
of $M_{12}$ and $\Gamma_{12}$ are convention dependent but their relative
phase is a physical observable. The physical parameters describing
$D^0-\bar D^0$ mixing are then conveniently expressed as 
\begin{equation}
x_{12} \equiv 2 \frac{|M^D_{12}|}{\Gamma^D}\,, \quad 
y_{12} \equiv \frac{|\Gamma^D_{12}|}{\Gamma^D}\,, \quad 
\phi_{12} \equiv \mathrm{arg} \bigg(\frac{M^D_{12}}{\Gamma^D_{12}}\bigg)\,,
\end{equation}
where $\Gamma_D$ is the average decay width of the neutral $D$ mesons.
The HFAG collaboration has performed a fit to these three parameters,  taking into account 
possible direct CP violating effects in the decay amplitudes. The resulting 
 95\% C.L.\ allowed ranges~\cite{HFAG},
\be
x_{12} \in [0.25,\, 0.99]\,\%\,, \qquad  y_{12} \in [0.59,\, 0.99]\,\%\,, \qquad 
\phi_{12} \in [-7.1^\circ,\, 15.8^\circ]\,,  \label{eq:expth}
\ee
are consistent with no CP violation in the  $D^0$--$\bar D^0$ transition amplitude. 

A detailed translation of these bounds into corresponding 
constraints on the coefficients of dimension-six $\Delta C=2$ effective operators, 
obtained under the assumption that non-standard contributions can at most saturate the above experimental bounds,
can be found in ref.~\cite{Gedalia:2009kh}. For later purposes, we report here some of the 
most significant constraints. Defining the $\Delta C=2$ effective Hamiltonian at the high scale
as 
\be
{\cal H}_{\rm eff}^{\Delta C=2}=\frac{1}{ (1~{\rm TeV})^2} \sum_{i}  z_i Q_i^{cu}+ {\rm H.c.}\,,
\label{eq:HeffDC2}
\ee
with 
\bea
&& Q_2^{cu} = \bar u^\alpha_R c^\alpha_L\bar
u^\beta_R c^\beta_L~,   \qquad 
Q_3^{cu} = \bar u^\alpha_R c^\beta_L\bar
u^\beta_R c^\alpha_L~,\no \\
&& Q_4^{cu}=\bar u^\alpha_R c^\alpha_L\bar
u^\beta_L c^\beta_R~, \qquad 
Q_5^{cu} = \bar u^\alpha_R c^\beta_L\bar
u^\beta_L c^\alpha_R~,
\eea
the bound on $|x_{12}|$  implies~\cite{Gedalia:2009kh}
\bea
&& |z_2| < 1.6\times10^{-7}~,\qquad 
|z_3| < 5.8\times10^{-7}~, \no \\
&& |z_4| < 5.6\times10^{-8}~,\qquad 
|z_5| < 1.6\times10^{-7}~,
\label{eq:DDbounds}
\eea 
while the bound on $|\phi_{12}|$ implies
$|{\rm Im}(z_i)| < 0.2 |z_i|^{\rm max}$, where $ |z_i|^{\rm max}$ are the numerical values shown in eq.~(\ref{eq:DDbounds}).

In general, the time-integrated CP asymmetry for neutral $D$ meson decays into 
a CP eigenstate $f$, defined in eq.~(\ref{eq:af_def}),  receives both 
direct and indirect CP-violating contributions. Expanding to first order 
in the CP-violating quantities we have~\cite{Grossman:2006jg,Bigi:2011re}
\begin{equation}
a_f = a^{\rm dir}_{f} + \frac{\langle t \rangle }{\tau} a^{\rm ind}~,
\label{eq:af_def2}
\end{equation}
with $a^{\rm dir}_{f} $ defined in eq.~(\ref{eq:Adirdef}) and where $a^{\rm ind}(x_{12},y_{12},\phi_{12} )$
is a universal term due to CP violation in the mixing amplitude or in 
the interference between mixing and decay.
By construction, the universal term cancels in the difference 
of two asymmetries into two different final states. This cancellation
is not exact in eq.~(\ref{LHCbnews}) because of different proper-time cuts 
in the two decay modes~\cite{lhcb}. This effect, as well as similar corrections 
in the previous measurements of  time-integrated CP asymmetries~\cite{Aaltonen:2011se, Staric:2008rx, Aubert:2007if}, 
have been taken into account by HFAG~\cite{HFAG} in obtaining the averages
reported in (\ref{eq:acpExp}) and (\ref{eq:expth}). 

\bigskip

\section{Other observables}
\label{sec:Other}

\subsection{CP violation in $B_{d,s}$ mixing} 
\label{sec:UT_analysis}

As pointed out by various authors (see e.g.~refs.~\cite{SL,Buras:2008nn,Altmannshofer:2009ne,Barbieri:2011ci,Lenz:2010gu,Bevan:2011zz}),  recent data on CKM fits show some tension. In particular, 
the predicted SM value of CP violation in $B_d$--$\bar B_d$ mixing (obtained removing the 
information on $S_{\psi K_S}$ from the global fits) 
and its direct determination via the 
time-dependent CP asymmetry in $B_d \to \psi K_S$ decays ($S_{\psi K_S}$)
are not in good agreement.
As we show in the following, this tension can be ameliorated 
in some of the  NP scenarios introduced to generate a 
sizable non-standard contribution to $\Delta a_{CP}$.
In other frameworks, detectable deviations from the SM are 
expected in the CP violating phase of $B_s$--$\bar B_s$ mixing,
measured via the time-dependent 
CP asymmetry  in  $B_s \to \psi \phi$  ($S_{\psi\phi}$).

In order to discuss these observables,  we start 
decomposing the $B_{d,s}$ mixing amplitudes as 
\begin{equation}
M^{q}_{12}=\left(M^{q}_{12}\right)_{\text{SM}} C_{B_q}e^{2 i\varphi_{B_q}}~,
\qquad (q=d,s)~.
\label{eq:M12}
\end{equation}
With this decomposition, the SM limit is recovered for $C_{B_q}=1$ and 
$\varphi_{B_q}=0$.  The $B_{s,d}$ mass differences and the CP asymmetries $S_{\psi K_S}$ 
and $S_{\psi\phi}$ assume the form
\bea
\Delta M_q &=& 2\left|M_{12}^q\right| = (\Delta M_q)_{\text{SM}}C_{B_q}~,
\label{eq:Delta_Mq} \\
S_{\psi K_S} &=& \sin( 2\beta + 2\varphi_{B_d} )~,\\
S_{\psi\phi} &=& \sin( 2|\beta_s| - 2\varphi_{B_s} )~,
\label{eq:CPV_Bq}
\eea
where the expressions of $S_{\psi K_S}$ and $S_{\psi\phi}$ are obtained under the 
assumption of negligible direct CP violation in the corresponding (tree-level, SM dominated) 
decay amplitudes. 

The phases $\beta$ and $\beta_s$ are defined by means of $V_{td}=|V_{td}|e^{-i\beta}$
and $V_{ts}=-|V_{ts}|e^{-i\beta_s}$. From global CKM fits based only on tree-level observables, 
or with arbitrary NP contributions to $\Delta F=2$ observables, it follows that~\cite{Bevan:2011zz}
\bea
  \sin(2\beta)_{\rm tree} &=& 0.775\pm 0.035~, 
  \label{eq:sin2btree} \\
  \sin(2\beta_s)_{\rm tree}  &=& 0.038\pm 0.003~.
\eea
These values have to be compared with the experimental 
determinations of the time-dependent  CP asymmetries~\cite{HFAG,LHCb:2011aa}.
\bea
  S_{\psi K_S}^{\rm exp} &=& 0.676\pm 0.020~,   \label{eq:sin2bexp} \\ 
  S_{\psi \phi(f_0)}^{\rm exp}  &=& - 0.03\pm 0.18~.
\eea
Direct and indirect determinations of $\sin(2\beta)$
differ by about $2.5 \sigma$. This tension can be eliminated by introducing 
a small new-physics contribution to the  $B_d$--$\bar B_d$ mixing
amplitude, such that  $\varphi_{B_d}\approx -5^\circ$.\footnote{Other mechanisms proposed to 
ameliorate the tension, introducing non-standard contributions to $\epsilon_K$ and/or $b\to u\ell\nu$
decays (see e.g.~ref.~\cite{Altmannshofer:2009ne}),  are not relevant for our following discussion.}
The determination of $\sin(2\beta_s)$ does not show any significant 
deviation form the SM expectation; however, the sizable error in 
$S_{\psi \phi}$
still allows for a new-physics correction of comparable size 
($|\varphi_{B_d}| \approx 5^\circ$--$10^\circ$).

As far as the moduli of the amplitudes are concerned, 
the SM predictions of $\Delta M_d$ and $\Delta M_s$ are both affected by 
$25-30\%$ errors (at the  $1\sigma$ level) and 
do not allow us to exclude stringent bounds.
The ratio of the two amplitudes is know 
to better accuracy ($\pm 13\%$ at the $1\sigma$ level, see table~\ref{tab:flavour}).

\subsection{Top FCNC}

As we will discuss later, concrete new-physics scenarios explaining the observed CP violation in $D$ decays generally
imply also large effects in FCNC top decays. We can parametrize the FCNC effects in the top sector in terms of the effective Lagrangian 
\begin{eqnarray}
-\mathcal{L}^\mathrm{eff} & = & \frac{g}{2 c_W}\, \bar q \gamma_\mu
\left( g^{qt}_{ZL} P_L + g^{qt}_{ZR} P_R \right) 
t Z^\mu 
  + \frac{e}{2m_t} \bar q
\left( g^{qt}_{\gamma L} P_L + g^{qt}_{\gamma R} P_R \right)
\sigma_{\mu \nu} t F^{\mu\nu}  \nonumber \\
&+& \frac{g_s}{2m_t} \bar q \left( g^{qt}_{g L} P_L + g^{qt}_{g R} P_R \right)
\sigma_{\mu \nu}  T^a t G^{a\mu\nu}  +\bar q 
\left(g^{qt}_{hL} P_L + g^{qt}_{hR} P_R \right) t H + \mathrm{h.c.} 
\label{ec:1}
\end{eqnarray}
With this notation, the top  FCNC decay widths are 
\begin{eqnarray}
\Gamma(t \to qZ) & = & \frac{\alpha_2}{32 c_W^2} |g^{qt}_{Z}|^2
\frac{m_t^3}{m_Z^2} \left( 1-\frac{m_Z^2}{m_t^2} \right)^2
\left( 1+2 \frac{m_Z^2}{m_t^2} \right) \,, \nonumber \\
\Gamma(t \to q \gamma) & = & \frac{\alpha}{4}
|g^{qt}_{\gamma}|^2 m_t \,, \nonumber \\
\Gamma(t \to q g) & = & \frac{\alpha_s}{3}
|g^{qt}_{\gamma}|^2 m_t \,, \nonumber \\
\Gamma(t \to q H) & = & \frac{m_t}{32\pi} |g^{qt}_{h}|^2
\left( 1-\frac{M_H^2}{m_t^2} \right)^2 \,,
\label{ec:3}
\end{eqnarray}
where $|g^{qt}_{X}|^2=(|g^{qt}_{XL}|^2+|g^{qt}_{XR}|^2)$ with $X= Z,\gamma,g,h$.

\section{Supersymmetry}
\label{sect:SUSY}

As a first example of explicit new-physics models  that can induce an enhanced chromomagnetic operator, we consider the supersymmetric 
extension of the SM with non-standard sources of flavor symmetry breaking.
In particular, we start by considering left-right flavor-breaking terms mixing the first two 
families of up-type squarks. Following usual notations, we call $\delta_{LR}$, $\delta_{LL}$, and $\delta_{RR}$
the ratios of off-diagonal terms in the squark squared-mass matrix (in the left-right, left-left, and right-right sectors, respectively) over the average squark squared mass, under the assumption that squarks are nearly degenerate in mass.

There is a fundamental reason why a left-right squark mixing is very suitable to explain the LHCb observation. Usually 
left-left or right-right squark flavor mixings are more constrained by $\Delta F =2$ processes  rather than $\Delta F =1$ 
transitions. This is because they give 
rise to corrections relative to the SM of the order of 
$\delta^2_{LL,RR}/(V_{ti}V_{tj}^*)^2 \times (m_W^2/{\tilde m}^2)$  in $\Delta F=2$ amplitudes 
-- where  $V_{ti}V_{tj}^*$ is the leading CKM factor -- and of the order of 
$\delta_{LL,RR}/(V_{ti}V_{tj}^*) \times (m_W^2/{\tilde m}^2)$ in $\Delta F=1$ amplitudes.
Thus if we assume that the supersymmetric contribution in a given $\Delta F =2$ observable does not exceed the SM one, we get 
the condition  
$\delta_{LL,RR}\sim V_{ti}V_{tj}^*~ {\tilde m}/m_W$. On the other hand, the corresponding bound from the  $\Delta F=1$  amplitude is parametrically weaker by an extra factor ${\tilde m}/m_W$.    For this reason, a left-left or right-right insertion is inadequate to explain the LHCb observation, as the constraints on $\delta^u_{LL}$ and $\delta^u_{RR}$ from $D^0-\bar{D}^0$ mixing are too strong.

The situation is reversed in the case of the contribution from left-right mixings. The unavoidable chiral suppression (proportional to the quark mass) hidden in the left-right mixing $\delta_{LR}$ becomes part of the structure of the dimension-six chromomagnetic operator, as defined in eq.~(\ref{eq:Q8def}). Since the quark mass participates in the hadronic matrix element, whose typical energy scale is the meson mass, it does not lead to any significant suppression factor. In other words, 
in $\Delta F =1$ processes the chiral suppression hidden in  $\delta_{LR}$ is compensated by the $v/m_c$ enhancement of the 
matrix element of the chromomagnetic operator.  On the other hand, contributions to $\Delta F =2$ transitions lead to four-fermion operators with coefficients proportional to $\delta_{LR}^2$. In this case, the double chiral insertion is not part of the operator and strongly suppresses $\Delta F =2$ processes. This means that, for the case under consideration in this paper, a mixing proportional to $\delta^u_{LR}$ gives a small effect to $D^0-\bar{D}^0$ mixing.
Moreover, the symmetry properties insure that an analogous double chiral suppression is needed to communicate the $\Delta C =1$ violation to a $\Delta S =1$ operator. For this reason, the left-right mixing in the up squarks can give a large contribution to CP violation in singly Cabibbo-suppressed $D$ decays, while producing small effects in $D^0-\bar{D}^0$ mixing and $\epsilon^\prime/\epsilon$.

We can now see explicitly how the mechanism work. 
In the supersymmetric framework,  
the leading contributions to $C_{8}$ and $\tilde C_{8}$ stem from loops involving up-squarks and gluinos. In the mass-insertion approximation, the expression for $C_{8}$
at the supersymmetric scale is
\be
\label{eq:C7_g}
C_{8}^{(\tilde g)} = -
\frac{\sqrt{2} \pi  \alpha_s \tilde m _{g}}{ G_F m_c}  \frac{  \left(\delta^{u}_{12}\right)_{LR} }{ {\tilde m_q^2}  }   ~g_{8}(x_{gq})~,  
\ee
where $\left(\delta^{u}_{12}\right)_{LR}$ denotes the left-right mixing in the first two generations of up-squarks (in the mass-eigenstate basis of up-type quarks) and $x_{gq}={\tilde m}_g^2/{\tilde m_q^2}$. The Wilson coefficient $\tilde C_{8}^{(\tilde g)}$ is obtained from $C_{8}^{(\tilde g)}$
via the replacement $\left(\delta^{u}_{12}\right)_{LR} \to   \left(\delta^{u}_{12}\right)_{RL} $,
and 
\begin{equation}
g_8(x) = \frac{11+x}{3(1-x)^3} + \frac{9+16x-x^2}{6(1-x)^4}\log{x}~,~~~~g_8(1) = -\frac{5}{36}.
\label{eq:g8}
\end{equation}
The enhancement factor $ {\tilde m}_g/m_c$ in eq.~(\ref{eq:C7_g}) is typically compensated by the chiral suppression (proportional to $m_c$) hidden inside the definition of $\left(\delta^{u}_{12}\right)_{LR}$. 

For later purposes, we report here also the results obtained in the case where the 1--2 transition arises from the mixing of the first two families with the third one. For near-degenerate squarks, 
we find
\be
C_{8}^{(\tilde g)} = -
\frac{\sqrt{2} \pi  \alpha_s \tilde m _{g}}{ G_F m_c}  \frac{   \left(\delta^{u}_{13}\right)_{LL}  \left(\delta^{u}_{33}\right)_{LR}  \left(\delta^{u}_{32}\right)_{RR} }{ {\tilde m_q^2}  }   ~F(x_{gq})~, 
\ee
\begin{equation}
F(x) = \frac{177+295x+7x^2+x^3}{36(1-x)^5} + \frac{9+50x+21x^2}{6(1-x)^6}\log{x}~,~~~~F(1) = -\frac{11}{360}.
\end{equation}
In the case of split families, in which only the third-generation squarks are light ($\tilde m_{q_{1,2}}^2 \gg \tilde m^2_{q_3}$), we find
\be
C_{8}^{(\tilde g)} = -
\frac{\sqrt{2} \pi  \alpha_s \tilde m _{g}}{ G_F m_c}  \frac{   \left(\delta^{u}_{13}\right)_{LL}  \left(\delta^{u}_{33}\right)_{LR}  \left(\delta^{u}_{32}\right)_{RR} }{ {\tilde m_{q_3}^2}  }   ~g_8(x_{gq})~, 
\ee
where the function $g_8(x)$ is given in eq.~(\ref{eq:g8}).
In the latter case $x_{gq_3}={\tilde m}_g^2/{\tilde m}_{q_3}^{2}$,
and  $\left(\delta^{u}_{i3}\right)_{LL,RR}$ are normalized to the heavy 
squarks masses (${\tilde m}_{q_{1,2}})$, while $\left(\delta^{u}_{33}\right)_{LR}$ 
is normalized to ${\tilde m}_{q_3}$.

The diagonal renormalization group evolution of the chromomagnetic operators down to the low
scales can be found, for instance, in ref.~\cite{Buras:1999da}.
To a good approximation, the main effect of the running 
is taken into account by evaluating the charm mass in eq.~(\ref{eq:C7_g}) at the low-energy
scale at which the hadronic matrix element is computed. Assuming, for illustrative purposes,
degenerate supersymmetric masses (${\tilde m_q}=\tilde m_{g}\equiv \tilde m$) and  $|(\delta^{u}_{12})_{LR}| \gg |(\delta^{u}_{12})_{RL}|$, we find
\be
\left|\Delta a^{\rm SUSY}_{CP} \right|   \approx   0.6\% \left(\frac{\left| \mathrm{Im} \left( \delta^{u}_{12}\right)_{LR} \right|}{10^{-3}}\right)
 \left(\frac{{\rm TeV}}{{\tilde m}}\right)~,
\label{eq:acpNPbis}
\ee
where we have used eq.~(\ref{blabla}) to estimate the matrix element of the chromomagnetic operator. This gives an uncertainty of order one in the coefficient in eq.~(\ref{eq:acpNPbis}). 

In a general supersymmetric framework, we expect the parametric relation
\be
\mathrm{Im} \left( \delta^{u}_{12}\right)_{LR}  \approx \frac{\mathrm{Im} (A)\ \theta_{12}\ m_c}{\tilde m} \approx  \left( \frac{ \mathrm{Im} (A)}{3}\right)\left(\frac{ \theta_{12}}{0.3}\right) \left(\frac{\rm TeV}{\tilde m}\right) 0.5 \times 10^{-3}~,
\label{eq:LR12dec}
\ee
where $A$ is the trilinear coupling and $\theta_{12}$ is a mixing angle between the first two generations of squarks. From eq.~(\ref{eq:acpNPbis}) we see that a large (and complex) trilinear coupling $A$, a Cabibbo-size mixing angle, and squarks with TeV masses give a value of $\mathrm{Im} \left( \delta^{u}_{12}\right)_{LR}$ in the correct ballpark to reproduce the required effect. Taking into account the large uncertainties involved in the evaluation of the matrix element, we conclude that a supersymmetric theory with left-right up-squark mixing can potentially explain the LHCb result.

To substantiate our conclusions, we need to check the consistency of $\left( \delta^{u}_{12}\right)_{LR}  \sim 10^{-3}$ with other measurements.
A double insertion of flavor-breaking terms $\left(\delta^{u}_{12}\right)_{LR}$ or  $\left(\delta^{u}_{12}\right)_{RL}$ 
in gluino up-squark loops leads to dimension-six $\Delta C=2$ effective operators contributing to $D-{\bar D}$ mixing.
Adopting the normalization of the $\Delta C=2$ effective Hamiltonian in eq.~(\ref{eq:HeffDC2}) 
we have 
\bea
     z_{2}^{(\tilde g)} &=& - \alpha_s^2 \left( \frac{{\rm TeV} }{   \tilde m_{q}} \right)^2 
        \left(\delta^{u}_{12}\right)^2_{RL} \frac{17}{18} x_{gq}\,f_6(x_{gq}) 
        \approx  - 5 \times 10^{-10}  \left( \frac{{\rm TeV} }{   m_{\tilde{q}}} \right)^2 \left[  \frac{ \left(\delta^{u}_{12}\right)_{RL}  }{1 \times 10^{-3} }
        \right]^2~,    \label{eq:z2} \\ 
      z_{4}^{(\tilde g)} &=& \alpha_s^2 \left( \frac{{\rm TeV} }{   \tilde m_{q}} \right)^2 
       \left(\delta^{u}_{12}\right)_{LR}\left(\delta^{u}_{12}\right)_{RL}  \frac{11}{18} \tilde{f}_6(x) 
       \approx  -2 \times 10^{-10}  \left( \frac{{\rm TeV} }{   m_{\tilde{q}}} \right)^2 \frac{ \left(\delta^{u}_{12}\right)_{LR} \left(\delta^{u}_{12}\right)_{RL}    }{ (1 \times 10^{-3})^2 }~, 
        \label{eq:z4}
\eea
$z_{3}^{(\tilde g)} = -(3/17)  z_{2}^{(\tilde g)}$ and $z_{5}^{(\tilde g)} = -(15/11)  z_{4}^{(\tilde g)}$, where 
$f_6(x)$ and $\tilde{f}_6(x)$ are the loop functions defined in ref.~\cite{Gabbiani:1996hi} such that $f_6(1)=1/20$ and $\tilde{f}_6(1)=-1/30$.  
The numerical values in eqs.~(\ref{eq:z2})--(\ref{eq:z4}) 
have been obtained in the limit of degenerate masses. 
As can be seen by comparing eqs.~(\ref{eq:z2})--(\ref{eq:z4}) with the bounds in eq.~(\ref{eq:DDbounds}),
values of  $\left(\delta^{u}_{12}\right)_{LR}$ or  $\left(\delta^{u}_{12}\right)_{RL}$  leading to 
$\Delta a_{CP} \approx 0.6\%$ are well below the current bounds 
from $D^0-\bar{D}^0$ mixing, in agreement with our general argument on left-right mixing effect given at the beginning of this section.

Through loops of charginos and up-squarks, $\left(\delta^{u}_{12}\right)_{LR}$ or  $\left(\delta^{u}_{12}\right)_{RL}$ can induce $\Delta S =1$ chromomagnetic or penguin operators. A simple inspection of the relevant Feynman diagrams shows that the effect will always be suppressed by $\left(\delta^{u}_{12}\right)_{LR}\left(\delta^{u}_{22}\right)_{RL}/{\tilde m}^2 \sim m_c^2/{\tilde m}^4$. Thus, the contribution to $\epsilon^\prime /\epsilon$ remains insignificant, even for $\left(\delta^{u}_{12}\right)_{LR}\sim 10^{-3}$.

\subsection{Disoriented $A$  terms}

The analysis we have just presented shows that an acceptable interpretation of the LHCb result can be given in terms of a supersymmetric theory with $\left(\delta^{u}_{12}\right)_{LR}\sim 10^{-3}$ and with small $\delta_{LL,RR}$. The possibility of the absence of flavor violation in the left-left and right-right sectors together with sizable effects in left-right transitions in not implausible in supersymmetric theories. This situation can be realized, for instance, when non-abelian flavor symmetries act on the $R$-invariant part of the supersymmetry-breaking terms, ensuring (total or partial) universality of soft masses, but are violated in the $R$-charged sector, allowing for general trilinear terms. Another possibility is that the pattern of supersymmetry breaking yields universal soft masses and general trilinear terms. Independently of the underlying explanation, the important point is that the separation between the properties of the soft terms and trilinear interactions of the first two generations is fairly robust. Indeed, the renormalization-group flow transfers flavor violation from the $R$-charged to the $R$-neutral sector, but it does so only through Yukawa interactions. Thus the effect is small for the first two generations relevant to the charm decays under consideration. 

While we can envisage scenarios in which flavor violation is restricted to the trilinear terms, it would be fairly unnatural to have this pattern only in the up sector, but not in the down sector. Therefore we generalize the structure of eq.~(\ref{eq:LR12dec}) to all squarks and take
\be
( \delta^{q}_{ij})_{LR}  \sim   \frac{ A   \theta^q_{ij}  m_{q_j} }{\tilde m}~~~~~q=u,d~,
\label{eq:LRgen}
\ee
where $\theta^q_{ij}$ are generic mixing angles. In table~\ref{tab:LRbounds} we summarize the present experimental constraints on 
$|\theta^q_{ij}|$ from flavor and/or CP violating processes. These results show that $\theta^q_{ij}$ can all be of order unity not only in the up, but also in the down sector, where experimental bounds on FCNC amplitudes are particularly stringent.
The tight limits on $( \delta^{d}_{ij} )_{LR}$ are naturally satisfied because of the 
smallness of down-type quark masses.
The only slightly problematic bounds in tab.~\ref{tab:LRbounds}, being significantly below unity, are those on $|\theta^{u,d}_{11}|$ coming from the neutron 
EDM~\cite{Hisano:2008hn}. The ansatz in eq.~(\ref{eq:LRgen}) could be compatible with the
EDM  constraints assuming non-maximal CP-violating phases.
In models in which the trilinear interactions follow the same flavor pattern of the Yukawa couplings (namely their entries are proportional to the corresponding entries of the Yukawa matrices up to coefficients of order one), we expect that $\theta^q_{ij}$ are roughly equal to the corresponding CKM angle $V_{ij}$. This pattern amply satisfies all bounds from flavor physics and predicts $\left(\delta^{u}_{12}\right)_{LR}\sim 10^{-3}$, in agreement with the LHCb observation.

\begin{table}[t]
\begin{center}
\renewcommand{\arraystretch}{1.2}
\begin{tabular}{|c||c|c|c|c|}
\hline
           &  $\theta^q_{11}$  & $\theta^q_{12}$  & $\theta^q_{13}$  &  $\theta^q_{23}$   \\ \hline\hline
q=d  &      $<  0.2$   &   $< 0.5$    &   $< 1$    &   --  \\ \hline
q=u      &  $< 0.2$      &   --  &  $<0.3$   &   $<1$   \\  \hline
\end{tabular} 
\renewcommand{\arraystretch}{1.0}
\caption{\label{tab:LRbounds}
Bounds on the moduli of the mixing angles $\theta^q_{ij}$, defined in eq.~(\ref{eq:LRgen}), 
assuming $A=3$, $\tilde m= 1$~TeV and maximal CP-violating phases. For $\theta^d_{ij}$ and $\theta^u_{11}$  the bounds
are derived from  gluino-mediated FCNCs or EDMs (see the bounds on the 
corresponding $\delta_{LR}$  in ref.~\cite{Isidori:2010kg,Hisano:2008hn}).
The bounds on $\theta^u_{i3}$ follow from the results in ref.~\cite{Isidori:2006qy}
on chargino-mediated FCNCs, assuming a degenerate supersymmetric spectrum. The
missing  entries have bounds exceeding unity.}
\end{center}
\end{table}

In conclusion, we have identified a specific structure of flavor violation in supersymmetric theories that can naturally explain the LHCb result, while satisfying all present constraints. In this scenario, which we call {\it disoriented $A$-terms}, the trilinear terms have the general form of eq.~(\ref{eq:LRgen}) both in the up and down sectors, while soft masses are (nearly) universal. This pattern can be obtained when the matrices of the up and down trilinear coupling constants follow the same hierarchical pattern as the corresponding Yukawa matrices but, in contrast with the usual minimal case, they do not respect exact proportionality. Since the trilinear and Yukawa matrices have the same transformation properties under the $U(3)^3$ flavor symmetry, it is plausible that, in certain setups, they follow the same hierarchical pattern, up to coefficients of order one in their individual entries.

Beside direct CP violation in the charm system, other signatures of this framework are electric dipole moments close to their 
upper bounds. Indeed the most stringent bounds in  table~\ref{tab:LRbounds} 
are those on $|\theta^{u,d}_{11}|$, set by the EDMs. 
Other potentially interesting observables are rare $B$ and $K$ decays induced by FCNC $Z$-penguins,
which in this framework are generated by chargino loops and are sensitive to $|\theta^{u}_{i3}|$. 
In clean processes such as $B_{s,d}\to \mu^+\mu^-$ and $K\to\pi\nu\bar\nu$
we can expect O($10\%-50\%$) deviations compared to the SM rates for  $|\theta^{u}_{i3}|$
close to their upper bounds~\cite{Isidori:2006qy}.

\subsection{Alignment models}

An interesting possibility to address the flavor problem in supersymmetry is provided by alignment mechanisms~\cite{Nir:1993mx,Leurer:1993gy},
which can naturally be implemented by means of Abelian symmetries~(see e.g.~ref.~\cite{Nir:2002ah} 
and references therein). The key feature of these mechanisms is to force an alignment between squark and quark mass matrices, in order to suppress dangerous FCNC effects, without requiring a degenerate squark spectrum, even among the first two families.

A general prediction of these models is a large left-handed mixing among the first two families
in the up sector, which seems to be a promising condition to get a large $\Delta a^{\rm SUSY}_{CP}$ according to eq.~(\ref{eq:acpNPbis}). In particular, the  left-handed mixing combined with a flavor-diagonal chirality breaking in the second generation yields an effective coupling relevant
for the chromomagnetic operator of the type
\be
\left(\delta^{u}_{21}\right)^{\rm eff}_{RL} = \left(\delta^{u}_{22}\right)_{RL}
\left(\delta^{u}_{21}\right)_{LL}~.
\label{eq:delta_alig}
\ee

The origin of the large $\left(\delta^{u}_{21}\right)_{LL}$ can be understood as follows.
The left-handed squark mass matrices in the basis where up or down quarks are diagonal
are related by $\tilde M^{(u)2}_{LL}= V \tilde M^{(d)2}_{LL} V^{\dagger}$, where $V$ is the 
CKM matrix. Expanding to first order in the Cabibbo angle ($\lambda=|V_{us}|$), we obtain
\begin{equation}
\label{qsa_approx}
(\tilde M^{(u)2}_{LL})_{21} \approx
(\tilde M^{(d)2}_{LL})_{21} + \lambda \left[(\tilde M^{(d)2}_{LL})_{22}-(\tilde M^{(d)2}_{LL})_{11}\right]~.
\end{equation}
Thus, even in the presence of a prefect alignment in the down sector (namely assuming $(M^{2(d)}_{LL})_{21}=0$), we find a sizable off-diagonal term in the up sector, as 
long as the left-handed squarks are non-degenerate,
\be
\label{defdel}
\left(\delta^{u}_{21}\right)_{LL} \approx \lambda ~{\Delta\tilde m^2_{21}\over\tilde m^2}~,
\ee
where $\Delta\tilde m^2_{21}$ is the square mass splitting between the first two generations
of left squarks. Similarly, one expects $(\delta^{u}_{32})_{LL} \sim |V_{cb}|$ and
$(\delta^{u}_{31})_{LL} \sim |V_{ub}|$, if the first two generations of squarks 
and the third one are not degenerate.

The assumption of an almost perfect alignment in the down sector allows us to evade all the stringent bounds from $K$ and $B$ physics.
One may worry that chargino-squark loops transfer the information of the flavor violation from the 
up sector to processes involving down-type FCNC. However, this is not the case because the chargino induced amplitudes are proportional to $V^\dagger \tilde M^{(u)2}_{LL} V= \tilde M^{(d)2}_{LL} $,
and thus they are diagonal in flavor. This result can be understood from general symmetry arguments. Let us consider $\tilde M^{2}_{LL}$ as a spurion of the approximate $U(3)^3$ quark flavor symmetry~\cite{MFV}. If $\tilde M^{2}_{LL}$ and the two quark Yukawa couplings ($Y_{u,d}$) are
the only sources of $U(3)^3$ breaking, and if $\tilde M^{2}_{LL}$ and $Y_d$ are diagonal in the
same basis, the only way to generate flavor-breaking effects in the down sector is by means of appropriate insertions of $Y_u$, which are strongly suppressed as in minimal flavor violation.

In spite of the weak constraints from the down sector, significant bounds on $\left(\delta^{u}_{21}\right)_{LL} $,
or equivalently on  $\Delta\tilde m^2_{21}$,  can be derived from $D$--$\bar D$ mixing.
According to the recent analysis in ref.~\cite{Gedalia:2009kh},  for squarks of at most
1~TeV, the mass splitting $\Delta\tilde m^2_{21}$ cannot exceed $15\%$ even in absence 
of new CP-violating phases.
As a result, we conclude that in alignment models  $|\left(\delta^{u}_{21}\right)_{LL}|$ does not exceed 3 $\times 10^{-2}$. Moreover, for TeV squarks, $(\delta^{u}_{22})_{RL} \approx Am_c/{\tilde m}$ cannot exceed about $10^{-3}$, or else the trilinear coupling $A$ would destabilize the vacuum.
As a result, from eqs.~(\ref{eq:delta_alig}) and (\ref{eq:acpNPbis}) we conclude  that in generic models of alignment $\Delta a^{\rm SUSY}_{CP}$ is predicted to be well below the central value of the recent
LHCb result. 

The only possibility to generate a large $\Delta a^{\rm SUSY}_{CP}$  in models 
of alignement, evading the bounds from $D$--$\bar D$ mixing, 
occurs if we assume that the  third generation of squarks is substantially lighter than the first two.
This is a specific case of what we call split family scenario and that will be discussed next.

\subsection{Split families}

The most severe suppression in the structure of $\left(\delta^{u}_{21}\right)^{\rm eff}_{RL}$ 
shown in eq.~(\ref{eq:LR12dec}) is the smallness of the charm mass, or the chirality flip 
in the second generation. This suppression can be partially avoided by generating the effective 
1-2 mixing through the coupling the first two generations to the third one, while taking advantage
of the large left-right mixing in the stop sector. This possibility is naturally realized in the supersymmetric framework with split families~\cite{io,altri}, where the first two generations of
squarks are substantially heavier than $\tilde t_{1,2}$ and  $\tilde b_{L}$,  the only squarks
required to be close to the electroweak scale by naturalness arguments. Originally formulated
in order to ameliorate the naturalness and flavor problems in supersymmetry, this framework is
further motivated at present by the absence of direct signals of supersymmetry at the LHC.
Indeed, while present LHC data exclude squarks  and gluinos below about 1 TeV in the case of a degenerate squark spectrum, the bounds on the stop-sbottom sector (and partially also on gluinos)
are much weaker in the case of a spectrum with split families.

Within this framework we can decompose the effective couplings relevant to $\Delta a^{\rm SUSY}_{CP}$ 
as follows
\be
\left(\delta^{u}_{12}\right)^{\rm eff}_{RL} =
\left(\delta^{u}_{13}\right)_{RR}\left(\delta^{u}_{33}\right)_{RL}
\left(\delta^{u}_{32}\right)_{LL}~,  \qquad 
\left(\delta^{u}_{12}\right)^{\rm eff}_{LR} =
\left(\delta^{u}_{13}\right)_{LL}\left(\delta^{u}_{33}\right)_{RL}
\left(\delta^{u}_{32}\right)_{RR}~.
\label{eq:12LReffSUSY}
\ee
This decomposition allows us to draw the following general considerations.
\begin{description}
\item[{\it LR mixing:}]
Since $\left(\delta^{u}_{33}\right)_{RL}$ in the stop sector can be approximately equal to one, it does 
not represent a significant suppression factor. Note that a Higgs mass
around 125 GeV,  as recently hinted by the LHC experiments~\cite{ATLAS,CMS}, naturally favors a large 
$A$ term if we want to keep the stop below 1 TeV. Thus the recent Higgs data support the assumption that $\left(\delta^{u}_{33}\right)_{RL}$ is of order unity.
In this limit, to generate sizable contributions to $\Delta a^{\rm SUSY}_{CP}$
we need $\left(\delta^{u}_{13}\right)_{LL} \left(\delta^{u}_{32}\right)_{RR}$ and/or 
$\left(\delta^{u}_{13}\right)_{RR} \left(\delta^{u}_{32}\right)_{LL}$ of $O(10^{-3})$. We also remark that, once we take $\left(\delta^{u}_{33}\right)_{RL}=O(1)$, there is no precise distinction between left and right sectors. Thus, from the phenomenological point of view, there is no difference between the case in which flavor mixings occur in the left-left and right-right sectors, as indicated by the decomposition in eq.~(\ref{eq:12LReffSUSY}), and the case in which flavor mixings originate from left-right trilinear terms of the type 3--1 and 3--2.
However, we find it useful to keep the decomposition~(\ref{eq:12LReffSUSY}) for illustrative purposes. 
\item[{\it RR mixing:}]
The mixing $\left(\delta^{u}_{3i}\right)_{RR}$, for $i=1,2$ in the up-type right-handed
sector is relatively unconstrained. The only significant bound comes from $D$--$\bar D$ 
mixing, which implies $|\left(\delta^{u}_{31}\right)_{RR} \times \left(\delta^{u}_{32}\right)_{RR}
|\simlt 10^{-2}$,
similarly to the limits on $|(\delta^{u}_{12})_{LL}|$ discussed in the case of alignment.
This bound refers to a squark mass of 1~TeV and the constraint scales linearly with the
stop mass. The limit from $D$--$\bar D$ mixing can be satisfied, for instance, for $\left(\delta^{u}_{32}\right)_{RR} = O(\lambda)$ and 
$\left(\delta^{u}_{31}\right)_{RR} = O(\lambda^2)$.
Another upper bound on flavor mixing is imposed by the condition that, in the case of a hierarchical squark spectrum and in the absence of tuning 
between large entries in the mass matrix leading to a small determinant, it is natural to expect
$|\left(\delta^{u}_{3i}\right)_{RR}| \lsim  \tilde m^2_{t_R}/\tilde m^2_{q_{i}}$ for $i=1,2$.

In this context it is worth to ask if right-handed mixing terms of similar size are allowed also
in the down sector.
In this case the constraints from $B_{d,s}$ mixing
imply $|\left(\delta^{d}_{31}\right)_{LL,RR}|  < \lambda^2$ and $|\left(\delta^{d}_{32}\right)_{LL,RR}|  < \lambda$ 
in the case of a split spectrum with $\tilde m_{q_3}=\tilde m_{g}=1$~TeV~\cite{ancorio} (see discussion on LL terms below).
So, the bounds on the $\left(\delta^{d}_{3i}\right)_{RR}$
are only one factor of the Cabibbo parameter $\lambda$ more stringent than the reference values for the $\left(\delta^{u}_{3i}\right)_{RR}$
assumed above.  It is also worth to stress that  in specific flavor models a breaking of flavor universality in the up-type right-handed 
sector larger than in the corresponding down-type  sector is not unlikely and may eventually be connected  to the large top-quark mass.
\item[{\it LL mixing:}]  The off-diagonal elements of the CKM matrix provide natural reference values 
for the mixing in the left-handed sector, namely  $|\left(\delta^{u}_{3i}\right)_{LL}| = O(|V_{ti}|$).
Even for such small mixing parameters  the effective couplings 
in eq.~(\ref{eq:12LReffSUSY}) can reach values of $10^{-3}$  
if the right-handed mixing terms are properly adjusted. In particular,  we can consider the following two options to explain the LHCb results:
\bea
\left(\delta^{u}_{32}\right)_{LL}=O(\lambda^2),\quad \left(\delta^{u}_{13}\right)_{RR} =  O(\lambda^2) \quad 
&\to& \quad \left(\delta^{u}_{12}\right)^{\rm eff}_{RL} = O(\lambda^4) = O( 10^{-3})~,  \no \\
\left(\delta^{u}_{13}\right)_{LL}=O(\lambda^3),\quad \left(\delta^{u}_{32}\right)_{RR} =  O(\lambda) \quad 
&\to& \quad \left(\delta^{u}_{12}\right)^{\rm eff}_{LR} = O(\lambda^4) = O( 10^{-3})~.
\label{pufpuf}
\eea
The two choices of mixing parameters in eq.~(\ref{pufpuf}) are mutually consistent and
thus both solutions can be simultaneously operative. The second solution can be realized 
in models of alignment, where 
$(\delta^{u}_{i3})_{LL} \sim |V_{ib}|$ and $(\delta^{u}_{i3})_{RR} \sim (m_{u_i}/m_{t})/|V_{ib}|$.
Note that, in the case of direct 1-2 mixing, the LHCb result could not be accounted for by
$LL$ or $RR$ mixing without getting into conflict with $D$--$\bar D$ mixing. In the case of
mixing through third generation this is instead possible, because it can be achieved through
much smaller mixing angles, taking advantage of the large chiral flip proportional to $m_t$.

As  far as direct experimental constraints are concerned, present data allow values of the left-handed mixing terms 
slightly exceeding the corresponding CKM factors~\cite{ancorio}. First of all, we note that   
$|\left(\delta^{d}_{3i}\right)_{LL}|$ can be smaller than $|\left(\delta^{u}_{3i}\right)_{LL}|$
assuming some alignment of the left-handed squark mass matrix to $Y_d$, as already discussed in the previous section. 
For $|\left(\delta^{u}_{3i}\right)_{LL}| \sim |\left(\delta^{d}_{3i}\right)_{LL}|$
the stronger bounds are derived from $B_d$ and $B_s$ meson mixing, for which we can write the following 
approximate formula
\be
M_{12}^q \approx  \left(M^q_{12}\right)^{\rm SM} 
\left[1+ \frac{  \left(\delta^{d}_{3q}\right)^2_{LL} }{ V_{tq}^2 } F_0~ \right]~,  \qquad 
F_0 \approx  \frac{1}{3} \left(\frac{g_s}{g} \right)^4 \frac{ m_W^2 }{ \tilde m^2_{q_3} } f_{0}(x_{gq_3})~,
\ee
where  $f_{0}(x)$ is a loop function normalized to 1 for $\tilde m_{q_3}=\tilde m_g$
(see ref.~\cite{Barbieri:2011ci}).
For $\tilde m_{q_3}=\tilde m_g=1$~TeV we find $|\left(\delta^{d}_{32}\right)_{LL}| < 0.2$
and $|\left(\delta^{d}_{31}\right)_{LL}| <  0.04$,  in agreement with ref.~\cite{ancorio}. 
For $|\left(\delta^{d}_{31}\right)_{LL}|$ close to its upper bound a non-negligible shift in the phase of $B_b$--$\bar B_d$ mixing
appears, offering the possibility to solve the discrepancy between $\sin(2\beta)_{\rm tree}$
and  $S_{\psi K_S}^{\rm exp}$ discussed in sect.~3.
\end{description}

\begin{figure*}[p]
\includegraphics[width=0.5\textwidth]{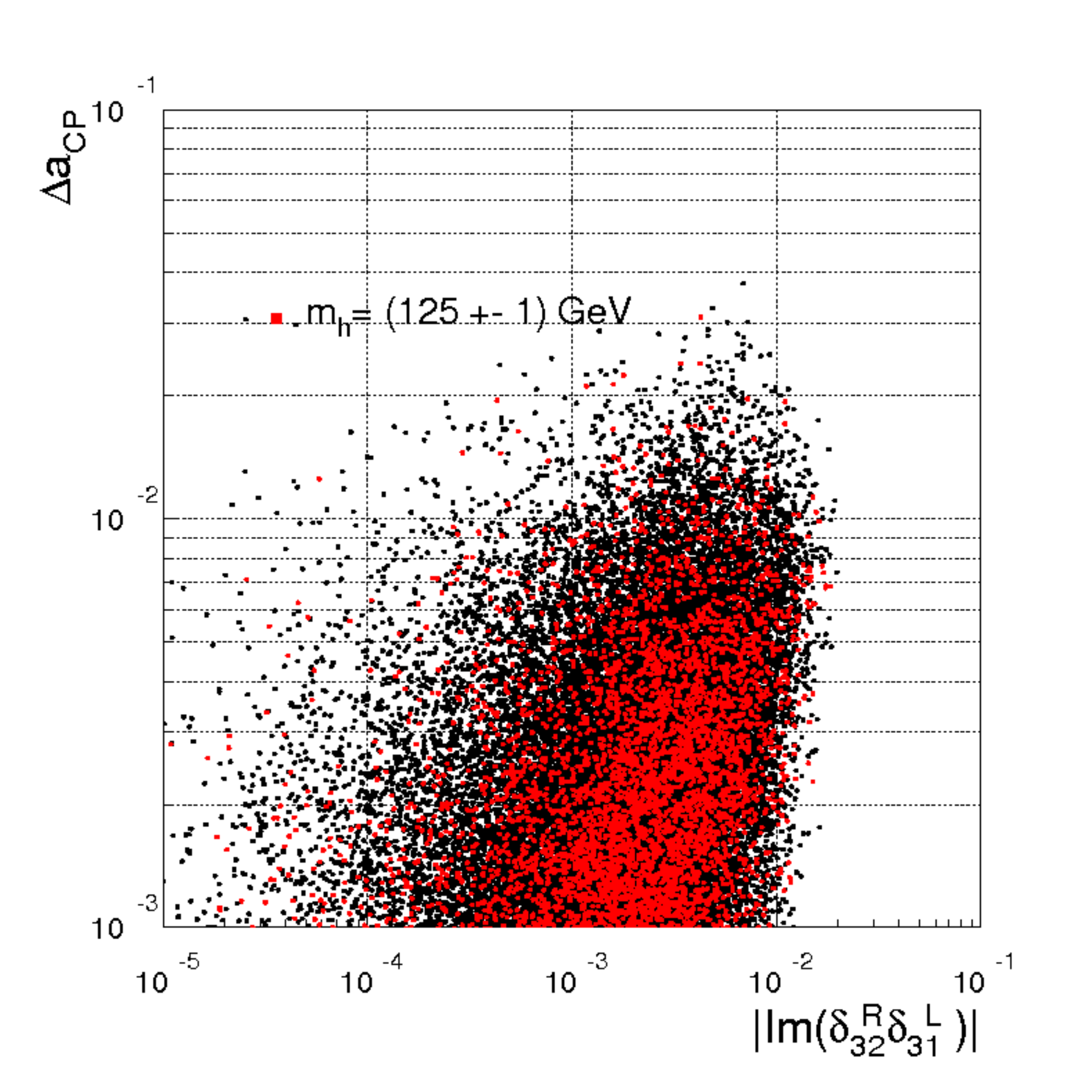}~~~
\includegraphics[width=0.5\textwidth]{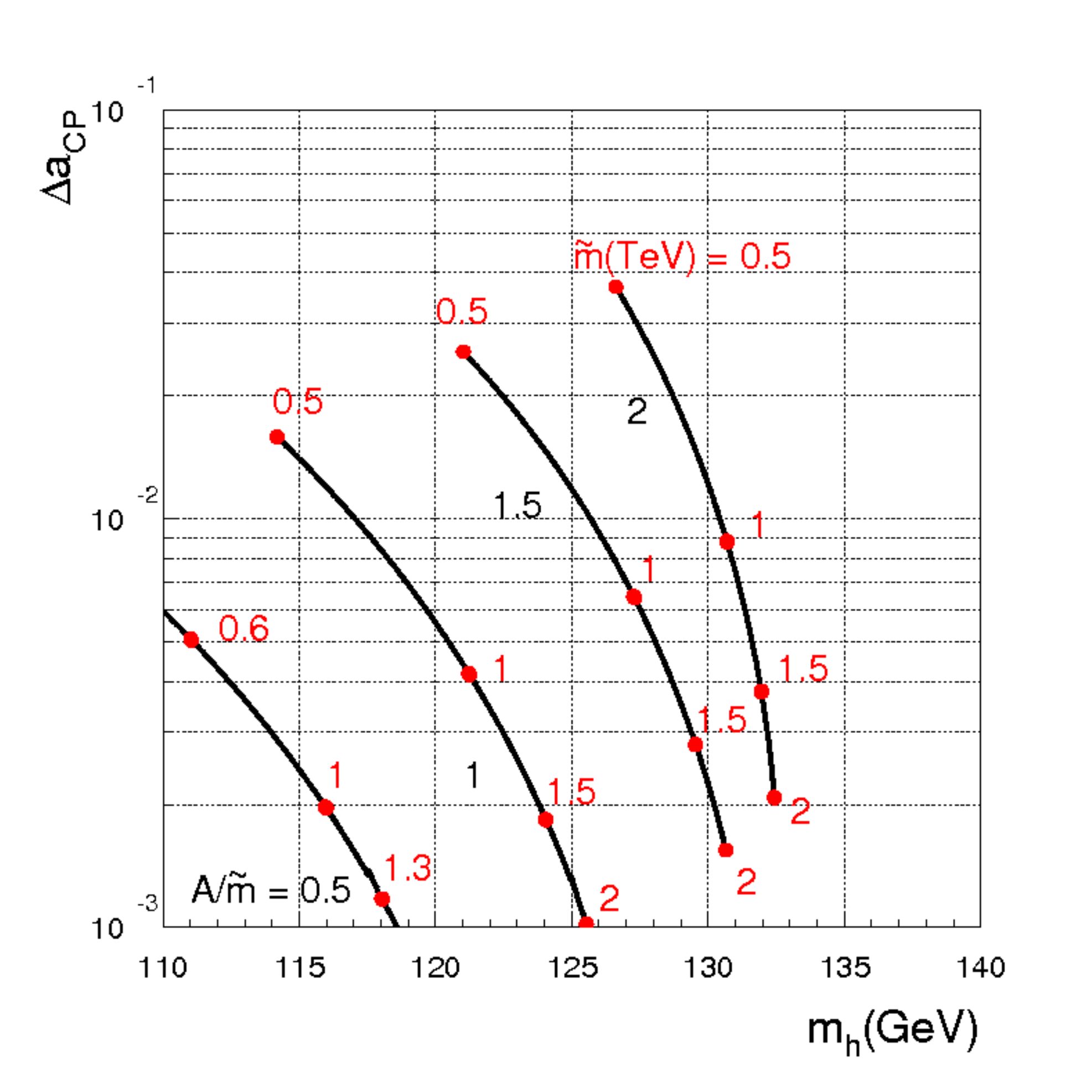}
\caption{
Left: $\Delta a^{\rm SUSY}_{CP}$ vs. $|{\rm Im}[(\delta^{u}_{32})_{RR}(\delta^{u}_{31})_{LL}]|$ 
for $0.5~{\rm TeV} \leq {\tilde m}, \tilde m_g \leq 2~{\rm TeV}$,
$\tan\beta=10$, $|A|\leq 3$ (see text for more details). The red points fulfill the condition
$m_h = (125 \pm 1)~$GeV. Right: $\Delta a^{\rm SUSY}_{CP}$ vs. $m_h$ for
$|{\rm Im} [(\delta^{u}_{32})_{RR}(\delta^{u}_{31})_{LL}]|=10^{-2}$, $\tilde m \leq 2$~TeV, 
and $A= 0.5, 1, 1.5, 2$.}
\label{fig:model_independent}
\end{figure*}

\begin{figure*}[p]
\includegraphics[width=0.5\textwidth]{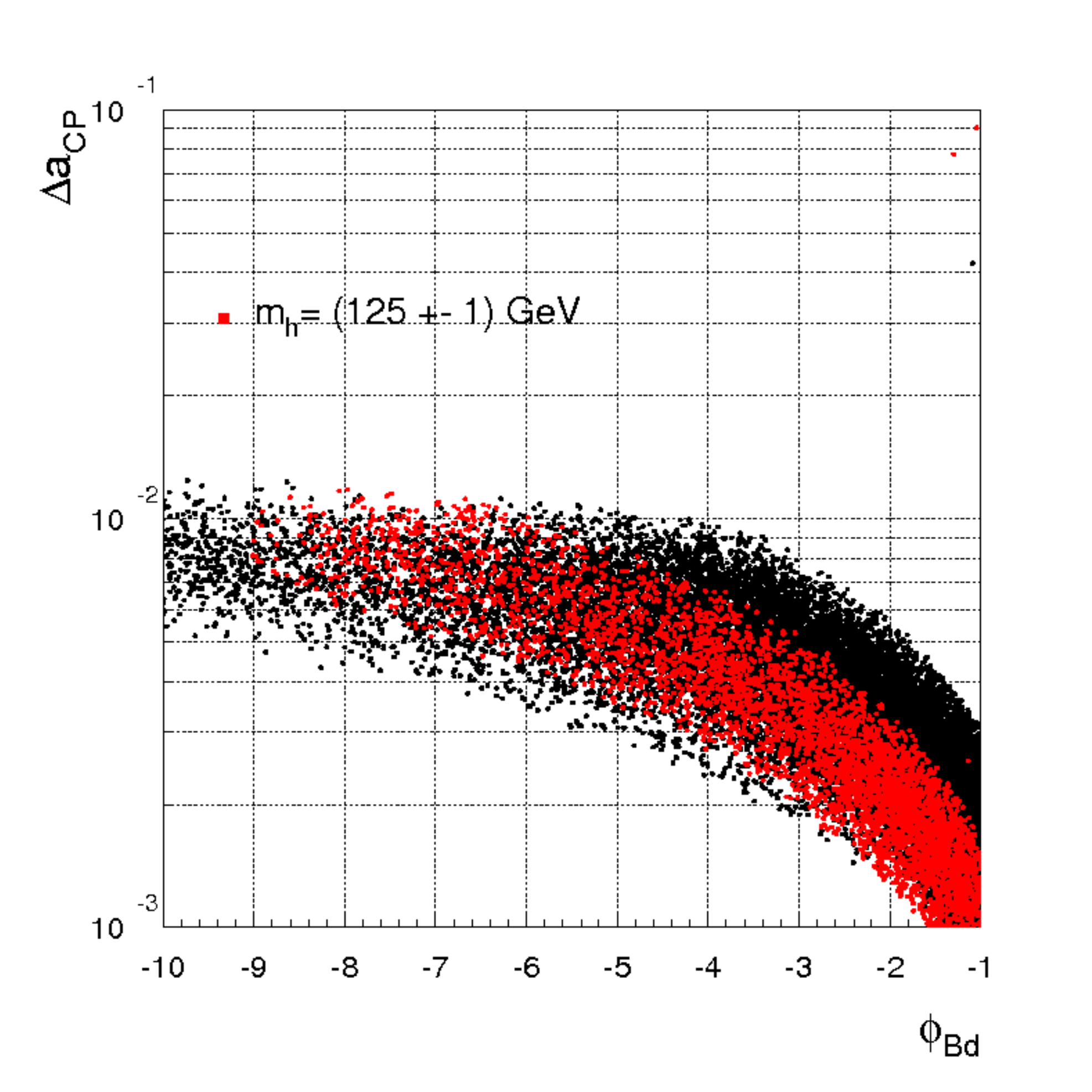}~~~
\includegraphics[width=0.5\textwidth]{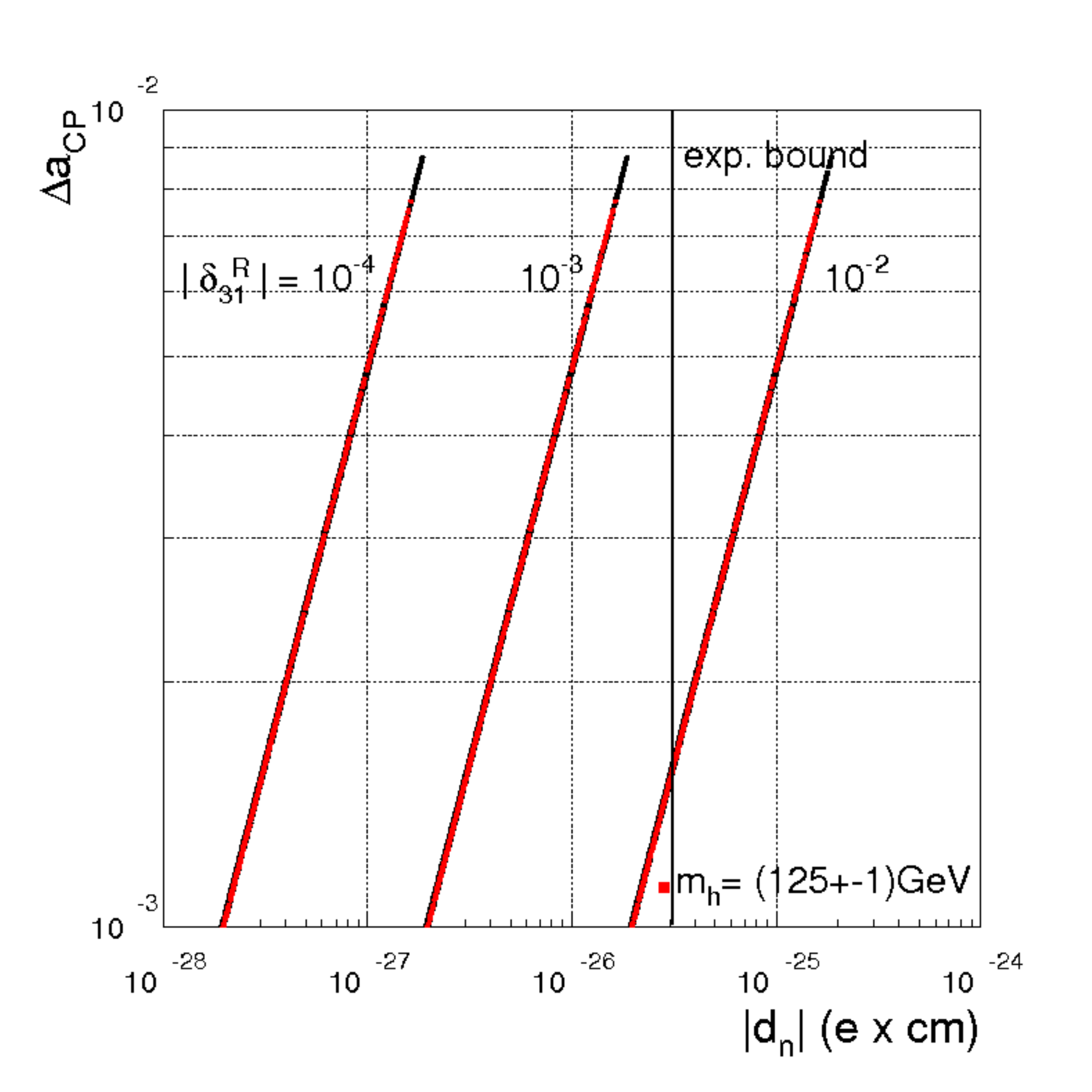}
\caption{Left: $\Delta a^{\rm SUSY}_{CP}$ vs.~$\varphi_{B_d}$ setting $(\delta^{u}_{32})_{RR} = 0.2$
and $\phi_{\delta^L_{31}} \in \pm(30^�,60^�)$, while varying $|(\delta^{d}_{31})_{LL}|<0.1$.
The other supersymmetric parameters are taken as in fig.~\ref{fig:model_independent}.
Right: $\Delta a^{\rm SUSY}_{CP}$ vs. $d_n$ assuming
$(\delta^{u}_{13})_{LL}= 10^{-2}$, $(\delta^{u}_{32})_{RR} = 0.2i$,
$|(\delta^{u}_{31})_{RR}| = 10^{-4},10^{-3},10^{-2}$ for the three lines and $\phi_{\delta^{R}_{31}}=30^{\circ}$.}
\label{fig:two}
\end{figure*}

After these general and qualitative considerations, we turn to a more detailed and quantitative analysis.
We perform a scan over the soft-breaking terms imposing the following conditions:
\begin{itemize}
\item
We set $ {\tilde m}\equiv \tilde m_{q_{3L}}=\tilde m_{t_R}$ and vary the parameters $\tilde m$, $\tilde m_g$ and $A$
in the following range: 500~GeV$ \leq \tilde m, \tilde m_{g} \leq 2$~TeV, $0 \leq |A/|\leq 3$.  
Flavor conserving parameters that do not play a direct role in $\Delta a_{CP}$ are not varied. In particular, 
we set  $\tan\beta=10$ and  $\mu=m_{H^\pm}=\tilde m_{\ell}=\tilde M_2=\tilde M_1=0.5~{\rm TeV}$
(where $\tilde M_{1,2}$ are the electroweak gaugino masses).
\item 
The $2\times 2$ blocks in the LL and RR squark mass matrices of the first two generations are assumed to be proportional 
to the identity matrix, with overall scale ${\tilde m^2}_{\rm heavy} = (5 {\tilde m})^2$. 
The 1--3 and 2--3 entries of the RR up-squark mass matrix (defined in the basis where the up quarks are diagonal)
and the LL squark mass matrix (defined in the basis where the down quarks are diagonal) are allowed to vary 
independently, with maximal size $\tilde m^2$. In this setup, the absence of special tunings in the squark mass matrix, together with the 
condition  ${\tilde m}_{\rm heavy}=5\tilde m$,  imply $|(\delta^{u,d}_{3i})_{LL}| , |(\delta^{u}_{3i})_{RR}| \lsim 0.1$. The $A$ terms satisfy exact proportionality.
\item
We evaluate all relevant FCNC amplitudes performing a complete diagonalization of the squark mass matrix --the mass-insertion language adopted so far was used only for illustrative purposes--  and impose the flavor constraints listed in table~\ref{tab:flavour}. We include leading QCD corrections
to all flavor observables. 
\item
The collider limits are applied by requiring $\tilde m_{g}> 500$~GeV and that the mass of the lightest stop is larger than 200~GeV, which roughly correspond to the present LHC bounds in the case of split families.
\end{itemize}
%
\begin{table}[t]
\addtolength{\arraycolsep}{3pt}
\renewcommand{\arraystretch}{1.2}
\centering
\begin{tabular}{|c|c|c|c|}
\hline
observable & experiment & SM prediction & exp./SM \\
\hline\hline
$|\epsilon_K|$ & $(2.229\pm 0.010) \times 10^{-3}$~\cite{Amsler:2008zzb} & $(1.90 \pm 0.26) \times 10^{-3}$~\cite{Brod:2010mj} & $1.17 \pm 0.16$ \\
\hline
$S_{\psi K_S}$ & $0.676\pm 0.020$~\cite{HFAG} & $0.775\pm 0.035$ & $0.87 \pm 0.05$ \\
\hline
$\Delta M_d$ & $(0.507\pm 0.005)$ ps$^{-1}$~\cite{HFAG} & $(0.51\pm 0.13)$ ps$^{-1}$ & $0.99 \pm 0.25$ \\
\hline
$\Delta M_s$ & $(17.77\pm 0.12)$ ps$^{-1}$~\cite{HFAG} & $(18.3\pm 5.1)$ ps$^{-1}$ & $0.97 \pm 0.27$ \\
\hline
$\Delta M_d/\Delta M_s$ & $(2.85 \pm 0.03) \times 10^{-2}$ & $(2.85 \pm 0.38) \times 10^{-2}$ & $1.00 \pm 0.13$\\
\hline
BR$(B\to X_s\gamma)$ & $(3.52\pm 0.25)\times 10^{-4}$ ~\cite{HFAG} &
$(3.15 \pm 0.23)\times 10^{-4}$~\cite{Misiak:2006zs} & $1.13  \pm  0.12$\\ 
\hline
$x_{12}$($D^0$--$\bar D^0$) &    $[0.25,\, 0.99]\%$   &   \multicolumn{2}{c|}{see sect.~\ref{sect:DDmix} } \\
$\phi_{12}$ ($D^0$--$\bar D^0$)  &   $ [-7.1^\circ,\, 15.8^\circ]$  &   \multicolumn{2}{c|}{  }     \\
\hline
\end{tabular}
\caption{\small
Experimental values and SM predictions for the most relevant observables used in our numerical analysis. The SM predictions for $\Delta F=2$ observables have been obtained by means of the CKM parameters determined in ref.~\cite{Bona:2005eu,Bevan:2011zz}
using tree-level observables only.}
\label{tab:flavour}
\end{table}
%
The results of the numerical scan  are illustrated in figs.~\ref{fig:model_independent} and \ref{fig:two}.
In the left panel of fig.~\ref{fig:model_independent}  we show $\Delta a^{\rm SUSY}_{CP}$ vs.~$|{\rm Im}[(\delta^{u}_{32})_{RR}(\delta^{u}_{31})_{LL})]|$. As can be seen, $\Delta a^{\rm SUSY}_{CP} \sim 0.6\%$ can be easily obtained for
$10^{-4}\lesssim |{\rm Im}[(\delta^{u}_{32})_{RR}(\delta^{u}_{31})_{LL})]| \lesssim 10^{-2}$,
in agreement with the qualitative discussion given above.
Interestingly enough, the condition  $m_h = (125 \pm 1)~$GeV on the Higgs mass is naturally
implemented in this sample (see red points). The underlying reason for this can be traced back
to the fact that both $\Delta a^{\rm SUSY}_{CP} > 10^{-3}$ and a large value of $m_h$ require
a sizable $A$-term in the stop sector, as already discussed.
To better show this correlation, on the right plot of fig.~\ref{fig:model_independent} we show $\Delta a_{CP}$ vs.~$m_h$ setting $|{\rm Im}[(\delta^{u}_{32})_{RR}(\delta^{u}_{31})_{LL})]|=10^{-2}$
and choosing $A= 0.5, 1, 1.5, 2$.
The main message we can read from this plot is that for
$A\sim 1$ and $\tilde{m}\sim$~TeV, $\Delta a^{\rm SUSY}_{CP}$ naturally 
lies in the per-cent range and $m_h \sim 125~$GeV, as hinted by recent data.


In the left panel of fig.~\ref{fig:two}  we show $\Delta a^{\rm SUSY}_{CP}$ vs. $\varphi_{B_d}$,
setting $(\delta^{u}_{32})_{RR} = 0.2$ and $\phi_{\delta_{31}^{L}} \in \pm(30^\circ,60^\circ)$,
with all the other parameters varied as in the general scan. As can be seen, an extra CP-violating
phase in the $B_d$ systems of order $\varphi_{B_d}\approx -5^\circ$, able to solve the current  discrepancy between $\sin(2\beta)_{\rm tree}$ and $S_{\psi K_S}^{\rm exp}$, can naturally be
obtained in the region of the parameter space also accounting for $\Delta a_{CP}\sim 0.6\%$.
Moreover, this region is also compatible with $m_h=(125 \pm 1)~$GeV (red points).
In the right panel of fig.~\ref{fig:two}  we show $\Delta a_{CP}$ vs.~the electric dipole moment
of the neutron, whose expectation within this framework is discussed in more detail below.

\subsubsection{Bounds from electric dipole moments}
\label{sect:edms}
The presence of new CP violating phases are expected to generate hadronic electric dipole moments. Gluino-squark loops, analogous to the one inducing the $\Delta C=1$ chromomagnetic operator, yield
an EDM ($d_u$) and a chromo-EDM ($d_u^c$) for the up quark.
In the limit of degenerate squark masses we have 
\be
 \left\{ \frac{d_{u}}{e},~d^{c}_{u}\right\} =
\frac{\alpha_s m_{\tilde g}}{4\pi \tilde{m}^{2}_{q}}\ f^{d_{u},d^{c}_{u}}(x_{gq_3})\ 
{\rm Im}\left[(\delta^{u}_{1i})_{LL}(\delta^{u}_{ii})_{LR}(\delta^{u}_{i1})_{RR}\right]\, .
 \label{Eq:edm_u_gluino}
\ee
Here $e$ is the electric charge, the index $i$ refers to the exchanged up-squark, and the loop functions (given explicitly in ref.~\cite{Hisano:2008hn}) are such that $f^{d_{u}}(1)= -8/135$, $f^{d^c_{u}}(1)=11/180$.
In the case of split squark masses the above expressions become
\be
 \left\{ \frac{d_{u}}{e},~d^{c}_{u}\right\} = - 
\frac{\alpha_s m_{\tilde g}}{2\pi \tilde{m}^{2}_{q_3}}\ f_3^{d_{u},d^{c}_{u}}(x_{gq})\ 
{\rm Im}\left[(\delta^{u}_{1i})_{LL}(\delta^{u}_{ii})_{LR}(\delta^{u}_{i1})_{RR}\right]\, ,
 \label{Eq:edm_u_gluino_split}
\ee
with
\be 
f_3^{d_{u}} (x) = - \frac{2(1+5x)}{9(1-x)^3} - \frac{4x(2+x)}{9(1-x)^4}\log{x}~, \qquad f_3^{d_{u}} (1) = -\frac{1}{27}~,
\ee
and $f_3^{d^{c}_{u}}(x)=g_8(x)$ with $g_8$ given in eq.~(\ref{eq:g8}).

Among the hadronic EDMs, the best constraints come from mercury and neutron EDMs ($d_{\rm Hg}$
and $d_n$ respectively). They can be expressed in terms of the up-quark EDM and chromo-EDM as~\cite{Pospelov:2005pr,Raidal:2008jk}
\begin{eqnarray}
\label{Eq:dn_odd}
d_n &\approx& (1\pm 0.5)\times \left(-0.35\,d_u + \,0.55\,e\,d^c_u\right)~, \\
\label{Eq:dHg}
d_{\rm Hg} &\sim& 7 \times 10^{-3} \,e\,d^c_u~.
\end{eqnarray}
For the leading contribution from stop exchange in the split-family case we find
\begin{eqnarray}
|d_n| \approx  \left|
\mathrm{Im}\left[(\delta^{u}_{13})_{LL}(\delta^{u}_{31})_{RR}\right] \right|
\left(\frac{{\rm TeV}}{{\tilde m}}\right)\,3\times 10^{-21}~{e\,\rm{cm}}~,\\
|d_{\rm Hg}| \approx  \left|
\mathrm{Im}\left[(\delta^{u}_{13})_{LL}(\delta^{u}_{31})_{RR}\right] \right|
\left(\frac{{\rm TeV}}{{\tilde m}}\right)\,2 \times 10^{-23}~{e\,\rm{cm}}~,
\label{eq:edmkkk}
\end{eqnarray}
to be compared with the current experimental bounds on $d_n$~\cite{Baker:2006ts}
and $d_{\rm Hg}$~\cite{Griffith:2009zz}:
\begin{eqnarray}
\label{edm_dn_exp}
|d_n| &<& 2.9 \times 10^{-26}~e\,\rm{cm}~(90\% \rm{CL})\,,\\
\label{edm_dHg_exp}
|d_{\rm Hg}| &<& 3.1 \times 10^{-29}~e\,\rm{cm}~(95\% \rm{CL})\,.
\end{eqnarray}
From the neutron EDM we obtain the bound
\be
\left| \mathrm{Im}\left[(\delta^{u}_{13})_{LL}(\delta^{u}_{31})_{RR}\right] \right|  <  10^{-5} \left( \frac{\tilde m}{\rm TeV} \right).
\label{resEDM}
\ee
The mercury EDM gives a bound which is even stronger, by about a factor of 10, but more sensitive
to nuclear uncertainties.

Although eq.~(\ref{resEDM}) involves two mixing angles that do not necessarily appear simultaneously
in the contribution to $\Delta a^{\rm SUSY}_{CP}$, it provides an important constraint on the interpretation of the LHCb result, once we make the assumption that left mixings are CKM-like.
For $(\delta^{u}_{13})_{LL} =O(\lambda^3)$, it implies $(\delta^{u}_{13})_{RR}\simlt 10^{-3}$,
which would eliminate the first solution in eq.~(\ref{pufpuf}) and require a strong hierarchy
among $|(\delta^{u}_{13})_{RR}|$ and $|(\delta^{u}_{23})_{RR}|$.

The potentially large correlation between $\Delta a_{CP}$ and $d_n$ is illustrated in the
right plot in fig.~\ref{fig:two}, where we show $\Delta a^{\rm SUSY}_{CP}$ vs. $d_n$ 
assuming $(\delta^{u}_{13})_{LL}= 10^{-2}$, $(\delta^{u}_{32})_{RR} = 0.2i$,
$|(\delta^{u}_{31})_{RR}| = 10^{-(2,3,4)}$  and $\phi_{\delta^{R}_{31}}=30^{\circ}$.
From the numerical analysis it turns out that
\be
\left|\Delta a^{\rm SUSY}_{CP} \right|
\approx
10^{-3} \times \left| \frac{d_n}{3 \times 10^{-26}} \right|
 \left| \frac{{\rm Im}\left(\delta^{u}_{32}\right)_{RR} }{0.2} \right|
\left| \frac{10^{-3}}{ {\rm Im}\left(\delta^{u}_{31}\right)_{RR} }\right|~.
\label{eq:acp_vs_dn}
\ee
In conclusion, we can have $\left|\Delta a^{\rm SUSY}_{CP} \right|\sim 0.6\%$ and, at the same time,
satisfy the EDM bounds. However, this requires a strong hierarchical structure in the off-diagonal terms of the RR up-squark mass matrix (or at least a sizable tuning of the corresponding CP-violating phases). Interestingly, this happens in specific models of alignment, where the off-diagonal terms
in the up right-handed sector are related to the up-quark masses by the relation
\be
(\delta^{u}_{ij})_{RR} \sim \frac{m_{u_i}/m_{u_j}}{|V_{ij}|}~.
\label{eq:deltaRalign}
\ee
Note that, even assuming such a strong hierarchical structure, 
the values of $d_n$ and $d_{\rm Hg}$ are expected to be very close to their present experimental bounds.

\subsubsection{Top and stop phenomenology}

The effective $\Delta C =1$ transition through third-generation squarks opens up the possibility of observing flavor violations in stop production and decays at the LHC. From the production point of view, the interesting process is $pp\to {\tilde t}^* {\tilde u}_i $, where ${\tilde u}_i ={\tilde u},{\tilde c}$. The rate for single ${\tilde u}_i$  production in association with a single stop is proportional to $(\delta^{u}_{i3})_{RR}^2$, since the mixings in the right-handed sector are larger then in the left sector. Besides testing the flavor structure, these processes allow us to extend the kinematical reach for the heavy squarks of the first two generations, although the production rates are typically small.

The flavor-violating stop decay is ${\tilde t} \to u_i \chi^0$, where $u_i =u,c$ and $\chi^0$ is the lightest neutralino. The width for the flavor-violating decay of any of the two stop mass eigenstates in units of the analogous decay into top is
\be
\frac{\Gamma (\tilde t \to c \chi^0)}{\Gamma (\tilde t \to t \chi^0)}=\left|(\delta^{u}_{i3})_{RR}\right|^2\left( 1-\frac{m_t^2}{{\tilde m}_t^2}\right)^{-2},
\ee
where we have neglected the neutralino mass with respect to the stop mass.

In order to benefit from larger production rates, it could be more interesting to consider gluino flavor-violating decays rather than the direct stop-pair production process. In models with split families, the gluino can decay only into third-generation squarks, ${\tilde g} \to {\bar t} {\tilde t}, ~{\bar b} {\tilde b}$. Once we include flavor violation, the decay ${\tilde g} \to {\bar u}_i {\tilde t}$ is also allowed, with a branching ratio proportional to the square of the corresponding mixing angle:
\be
\frac{\Gamma (\tilde g \to \tilde t  u_i)}{\Gamma (\tilde g\to \tilde t t)}=\left|(\delta^{u}_{i3})_{RR}\right|^2\left[1
+ O\left(\frac{m_t}{\tilde m_g}\right) \right]~.
\ee
If the gluino is not too heavy, the large number of events collected at the LHC could allow for an important test of this  flavor-violating mode.

\medskip

Having introduced effective couplings between the third and the first two generations in the up sector, 
in this framework we also have a natural link with rare flavor-violating top decays.
However, we have
explicitly checked that all the branching ratios for the relevant FCNC top decays
 lie below the $10^{-6}$
level, and thus are beyond the reach of near-future facilities (see e.g.~ref.~\cite{mele} and references therein).
Indeed, the loop suppression and the decoupling factor imply
\be
{\rm BR}(t\to qX) \sim 
\left(\frac{\alpha}{4\pi}\right)^2
\left(\frac{m_W}{m_{\rm SUSY}}\right)^4
|\delta^{u}_{3q}|^2
\ee
where $m_{\rm SUSY}= {\rm max}(m_{\tilde g},m_{\tilde t})$ for $X =\gamma,g,Z$
and $m_{\rm SUSY}= m_A$ for $X=h$. Therefore, even for maximal mixing angles,
i.e. $\delta^{u}_{3q}\sim 1$ and $m_{\rm SUSY}\gtrsim 3 m_W$, it turns out
that ${\rm BR}(t\to qX) \lesssim 10^{-6}$.

\section{Other  new-physics scenarios}
\label{sect:nonSUSY}

\subsection{New-physics scenarios with Z-mediated FCNC}
\label{sect:Zmediated}

Effective FCNC couplings of the $Z$ boson to SM quarks, or between quarks and heavier fermions, 
can appear in several new-physics frameworks~\cite{Nir:1990yq,Buchalla:2000sk,Langacker:2000ju}.
Prominent examples are the SM with non-sequential generations of quarks, models
with an extra $U(1)$ symmetry~\cite{Langacker:2008yv}
or models with extra vector-like doublets and singlets~\cite{Branco:1999fs}.

Irrespective of the underlying dynamics, 
we introduce the following effective Lagrangian to describe the FCNC couplings of the $Z$-boson to fermions
\be
\mathcal{L}^{Z-{\rm FCNC}}_{\rm eff} =
-\frac{g}{2 \cos \theta_W}
{\bar F}_i\gamma^\mu
\left[ (g^{Z}_{L})_{ij} \, P_L + (g^{Z}_{R})_{ij} \, P_R \right] q_j ~ Z_\mu
+ \text{ h.c.}\,,
\label{eq:eff_lagr_Z}
\ee
where $g$ is the $SU(2)_L$ gauge coupling, and 
$F$ can be either a SM quark ($F=q$) or some heavier non-standard fermion.
In the following we focus on the $F=q$ case.
However, many of the results can easily be generalized to the case 
where $F$ is some heavier state. Moreover, our discussion of flavor-violating interactions of the $Z$ boson can be extended in straightforward way to the case of new gauge bosons $Z^\prime$. 

If $F$ is a SM fermion, the effective Lagrangian in eq.~(\ref{eq:eff_lagr_Z})
breaks explicitly the electroweak symmetry. It is therefore natural to normalize 
the effective couplings $(g^{Z}_{FL})_{ij}$ and $(g^{Z}_{FL})_{ij}$ as follows
\be
(g^{Z}_{L})_{ij} = \frac{v^2}{M^{2}_{\rm NP}} (\lambda^Z_L)_{ij}\, ,
\qquad
(g^{Z}_{R})_{ij} = \frac{v^2}{M^{2}_{\rm NP}} (\lambda^Z_R)_{ij}\, ,
\ee
where $v$ is the SM Higgs vacuum expectation value ($v= 246$~GeV),
$M_{\rm NP}$ is the effective scale of the new dynamics generating the FCNC couplings,
and  $(\lambda^Z_{L,R})_{ij}$ are dimensionless flavor off-diagonal couplings.

The  chromomagnetic operator is generated at the one-loop level, with   
leading contribution from $Z$--top exchange diagrams (unless the corresponding couplings are
strongly suppressed). The one-loop expression for $C^{Z}_{8}$ is
\be
C^{Z}_{8} = \frac{m_t}{m_c} \, (g^{Z}_{L})^*_{ut}(g^{Z}_{R})_{ct}\, h_{8}(x_{tZ})\, ,
\label{eq:C8Z}
\ee
where $x_{tZ}=m^2_t/m^2_Z$ and 
\be
h_{8}(x)=\dfrac{4+x+x^2}{8(1-x)^2}+
\dfrac{3x\log{x}}{4(1-x)^3}\,.
\label{eq:h8}
\ee 
As usual, ${\tilde C}^{Z}_{8}$ is obtained from  ${C}^{Z}_{8}$ via $L\leftrightarrow R$.
For completeness, we note that at the same order also FCNC magnetic-dipole operators 
are generated\footnote{The FCNC magnetic-dipole operators 
$Q_{7}$ and $\tilde Q_7$ are defined as in (\ref{eq:Q8def})
with $T^a g_s G_a^{\mu\nu} \to eF^{\mu\nu}$.}
 with effective couplings $C^{Z}_{7} = Q_t C^{Z}_{8}$ and 
$\tilde C^{Z}_{7} = Q_t \tilde C^{Z}_{8}$, where $Q_t = Q_u = +2/3$. Using the results in 
sect.~\ref{sec:DCPC}, and considering only the contribution of ${C}^{Z}_{8}$,
we then find 
\be
\left| \Delta a^{Z-{\rm FCNC}}_{CP} \right| \approx 
0.6\%~\left| \frac{\mathrm{Im}\left[(g^{Z}_{L})^*_{ut}(g^{Z}_{R})_{ct}\right] }{2\times 10^{-4}}\right| 
\approx 0.6\%~\left| \frac{ \mathrm{Im}\left[(\lambda^{Z}_{L})^*_{ut}(\lambda^{Z}_{R})_{ct}\right]  }{5 \times 10^{-2} } \right|
\left(\frac{\rm 1~TeV}{M_{\rm NP}}\right)^4~. 
\label{eq:acpZNP}
\ee
As can be seen, the required value of $\Delta a_{CP}$ can be generated only if 
the effective scale $M_{\rm NP}$ 
is at most around 1 TeV and the flavor-violating couplings for the top quark
are large. Such a situation can occur in models where the top quark is a
composite or a partially-composite state of some new strongly interacting dynamics at the 
TeV scale. 

\subsubsection{Low-energy constraints}
Strong constraints on the $(g^{Z}_{L,R})_{ij}$ effective couplings for up-type quarks 
arise from $D^0-\bar{D}^0$ mixing: from tree-level $Z$ exchange diagrams we get $|(g^{Z}_{L,R})_{uc}|< 2\times 10^{-4}$
and  $|(g^{Z}_{L})_{uc}(g^{Z}_{R})_{uc}|< 0.5 \times 10^{-8}$. However, much weaker 
constraints are derived 
on the effective couplings involving the top, appearing in eq.~(\ref{eq:acpZNP}), 
since they contribute to  $D^0-\bar{D}^0$ mixing only at the one-loop level. 
Similarly to the case of supersymmetry with split families,  if we assume that the transition between the first two generations 
is  induced only as a result of  1--3 and 2--3 mixings, 
then the couplings necessary to generate
$|\Delta a_{CP}| \approx 10^{-2}~$ lead to effects in $D^0-\bar{D}^0$ mixing well
below the current experimental bounds.

Even if we set to zero the FCNC $Z$ couplings in the down sector at the scale $M_{\rm NP}$, 
in the left-handed sector they are induced at the one-loop level by $W$-up-quark loops.
In particular, the leading-log contributions to the down-type induced couplings are
\begin{equation}
(g^{Z}_{L})_{d_i d_j}  =  \frac{1}{ 8 \pi^2 } \frac{m_t^2}{v^2}  {\rm ln}\frac{m_t^2}{M_{\rm NP}^{2}}
\sum_q (g^{Z}_{L})_{tq}  V_{ti}^*V_{qj}
\label{eq:Zbb}
\end{equation}
where $V_{ij}$ is the CKM matrix. Similarly, one-loop Yukawa interactions leads to corrections to the 
down-type couplings of the type $(\delta g^{Z}_{L})_{d_i d_j}  \sim  (1/16 \pi^2 ) \sum_q  (m_t m_q/v^2) (g^{Z}_{R})_{tq}  V_{ti}^*V_{qj}$,
which are of comparable size if $(g^{Z}_{R})_{tq}/(g^{Z}_{L})_{tq} \sim 10^{2}$.
The  $(g^{Z}_{L})_{d_i d_j} $
couplings, in turn, are
severely constrained by down-type $\Delta F=2$ amplitudes and rare FCNC leptonic 
decays of $B$ and $K$ mesons (see e.g.~ref.~\cite{Buchalla:2000sk}). In particular,  
for $B_{s,d}$ mixing and the rare $B_{s,d}\to \mu^+\mu^-$ decays we have 
\bea
\frac{ M_{12}^q }{ \left(M^q_{12}\right)^{\rm SM} } &=&  1+ \frac{4\pi^2 }{\sqrt{2} G_F m^2_W S_0}
\left[  \frac{(g^{Z}_{L})_{bq} }{ V_{tq}V^{*}_{tb} } \right]^2
\approx 1+  \left[ \frac{(g^{Z}_{L})_{qb} }{0.08 V_{tq} } \right]^2 \,,
\label{eq:DF2_Z}  \\
\frac{\Gamma(B_{q} \to \mu^+ \mu^-)}{\Gamma(B_{q} \to \mu^+ \mu^-)_{\rm SM}} &=&
\left|  1 -  \frac{ \pi^2 }{\sqrt{2} G_F m^2_W Y_0}
 \frac{(g^{Z}_{L})_{bq} }{ V_{tq}V^{*}_{tb} } \right|^2 
 \approx \left|  1 -   \frac{(g^{Z}_{L})_{qb} }{0.01 V_{tq} } \right|^2 \,,
\eea
where $S_0\approx 2.3$ and $Y_0 \approx 1$ are the corresponding SM loop functions.
The strongest constraint follows from $B_{d}$ mixing, which implies $|(g^{Z}_{L})_{qb}|< 4\times 10^{-4}$.
Taking into account eq.~(\ref{eq:Zbb}), this condition is fulfilled assuming $|(g^{Z}_{L})_{tu}| < 2\times 10^{-2}$,
which does not prevent sizable contributions to $\Delta a_{CP}$ according to eq.~(\ref{eq:acpZNP}).
Moreover, close to the upper bound on $|(g^{Z}_{L})_{tu}|$, it
 is possible to generate a contribution to $B_{d}$ mixing that decreases the tension in the CKM fits.

The presence of new CP violating phases in the couplings $(g^{Z}_{L,R})_{ij}$ are
expected to generate also hadronic EDMs, with a strong correlation to $\Delta a_{CP}$
as already seen in the case of supersymmetry.
In particular, we find 
\be
d^{c}_{u} =
-\frac{\sqrt{2}G_F}{4\pi^2} m_t \,
{\rm Im} \left[ (g^{Z}_{L})^*_{ut} (g^{Z}_{R})_{ut} \right]\, h_{8}(x_{tZ})~,
\qquad
\frac{d_{u}}{e}=Q_u d^{c}_{u}~,
\label{eq:edm_Z}
\ee
which, according to eq.~(\ref{Eq:dn_odd}), leads to 
\be
|d_n| \approx 3  \times 10^{-26}~
\left| \frac{ {\rm Im} \left[ (g^{Z}_{L})^*_{ut} (g^{Z}_{R})_{ut} \right] }{ 2\times 10^{-7}} \right|
~{e\,\rm{cm}}\,.
\label{eq:dZNP}
\ee
Comparing the above result with eq.~(\ref{eq:acpZNP}) we find that, similarly to the case of supersymmetry, a contribution to $\Delta a_{CP}$ at the per-cent level is allowed only if there exists a strong hierarchy among the $(g^{Z}_{L,R})_{tq}$ couplings. For instance, assuming 
$(g^{Z}_{L})_{ut}$ to be real, $\Delta a^{Z-{\rm FCNC}}_{CP}=O(10^{-2})$ and the EMDs bounds are satisfied only if ${\rm Im}(g^{Z}_{R})_{ut}/{\rm Im}(g^{Z}_{R})_{ct}\lesssim 10^{-3}$.

\subsubsection{Implications for top-quark physics}
\begin{figure*}[t]
\includegraphics[width=0.5\textwidth]{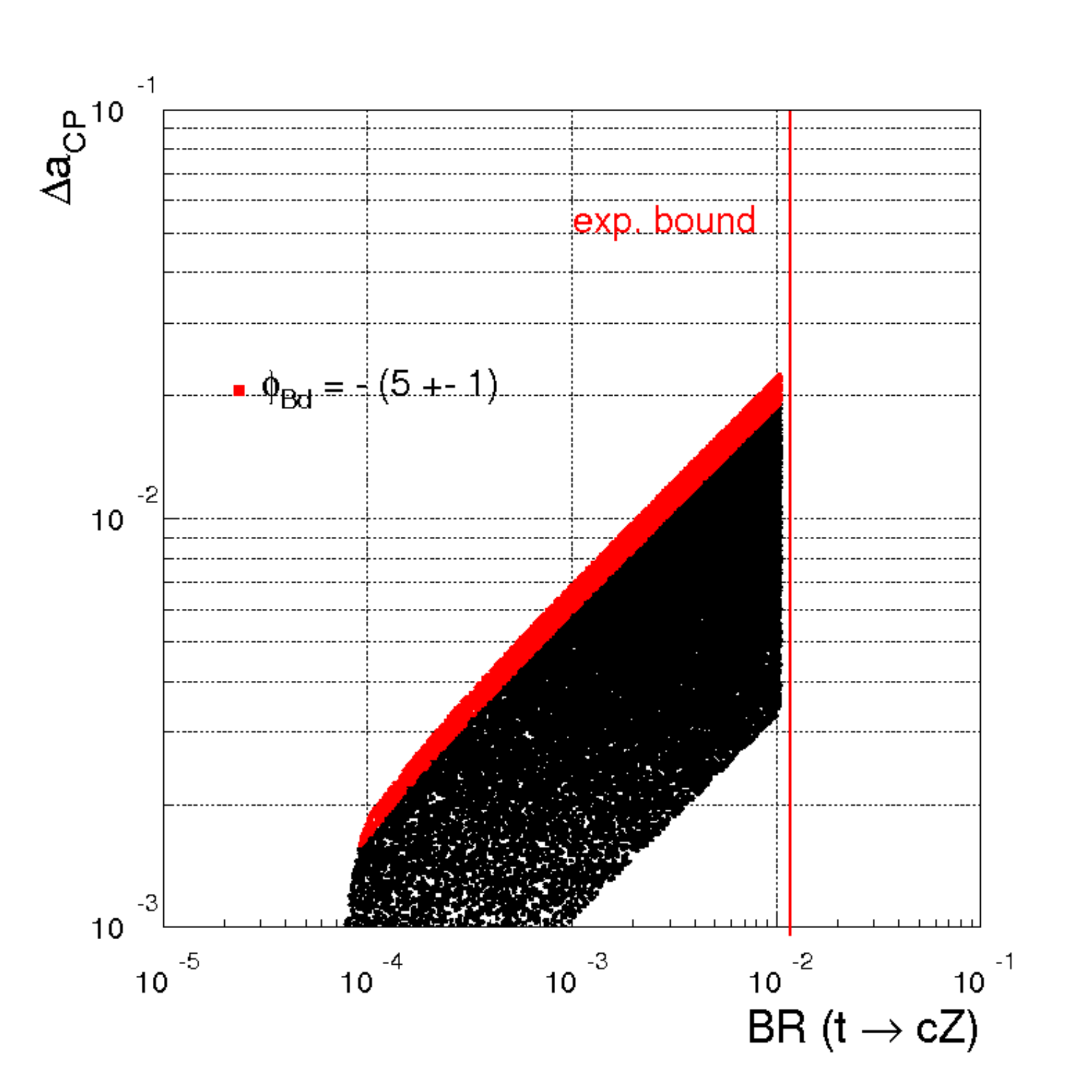}~~~
\includegraphics[width=0.5\textwidth]{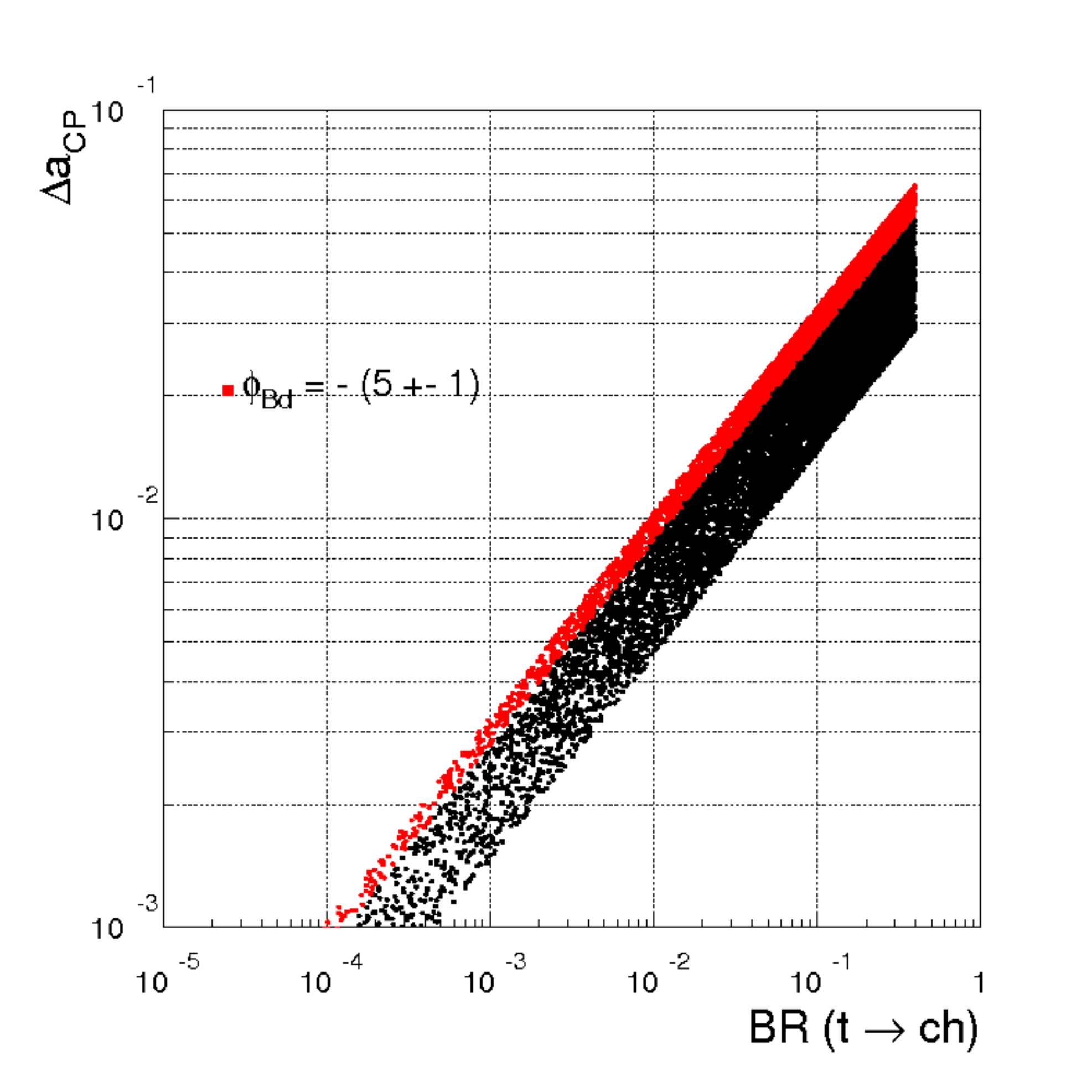}
\caption{
Left: ${\rm BR}(t\to cZ)$ vs.~$\Delta a^{Z-{\rm FCNC}}_{CP}$.
Right: ${\rm BR}(t\to c h)$ vs.~$\Delta a^{h-{\rm FCNC}}_{CP}$.
The plots have been obtained by means of the scan: $|(g^X_L)_{ut}|>10^{-3}$, $|(g^X_{R})_{ct}|>10^{-2}$, where $X=Z,h$, with arg$[(g^X_L)_{ut}]=\pm\pi/4$
and arg$[(g^X_R)_{ct}]=0$.
The points in the red regions solve the tension in the CKM fits through
a non-standard phase in $B_d$--${\bar B}_d$ mixing, assuming for the
corresponding down-type coupling $(g^X_L)_{db}= 5\times 10^{-2} (g^X_L)_{ut}$.}
\label{fig:model_topFCNC}
\end{figure*}

In the NP scenarios with Z-mediated FCNCs, the most interesting FCNC processes
in the top sector are $t\to cZ$ and $t\to uZ$, which arise at the
tree level.   In particular, we find that
\be
\mathrm{Br}(t \to cZ) \approx 0.7\times 10^{-2}
\left| \frac{ (g^{Z}_{R})_{tc} }{10^{-1}}\right|^2\,,
\ee
which is within the reach of the LHC for the values of $(g^{Z}_{R})_{tc}$
relevant to $\Delta a^{Z-{\rm FCNC}}_{CP}$, as illustrated in fig.~\ref{fig:model_topFCNC}.
Actually, the present
ATLAS bound $\mathrm{Br}(t \to qZ) < 1.1\%$~\cite{ATLAStopFCNC} already provides a
significant constraint on the model:   this constraint turns out to be slightly more stringent 
that the bound on $|(g^{Z}_{L})_{tq}|$ posed by electroweak precision observables 
 (in particular by the correction to the $\rho$ parameter).
On the other hand, the branching ratios of loop induced processes
such as $t\to q\gamma$, $t\to q g$ and $t\to q h$ are are well below the 
$10^{-6}$ level, and thus far from the experimental reach in the near future.

Finally, it is worth to mention that a non-vanishing ${\rm Im}(g^{Z}_{R})_{ut}$ could also
contribute to the forward-backward  asymmetry in $t\bar{t}$ production $(A^{t\bar t}_{FB}$), 
by means of a $t$-channel exchange of the $Z$ boson. This effect has been discussed in the recent 
literature~\cite{Jung:2009jz}, 
given the sizable discrepancy between data and SM predictions in
$A^{t\bar t}_{FB}$ observed at the Tevatron~\cite{Aaltonen:2011kc}.
Using the results of ref.~\cite{Jung:2009jz}, 
we find that the induced effect for the reference values of ${\rm Im}(g^{Z}_{R})_{ut}$ relevant 
to $\Delta a_{CP}$ is too small to explain the current $A^{t\bar t}_{FB}$ anomaly. 
Moreover, in the meanwhile the model with a single $t$-channel $Z$ exchange 
(or a generic $Z^\prime$ boson from $U(1)$ symmetries)
has been ruled as a possible solution to the  $A^{t\bar t}_{FB}$ anomaly 
because of the excessive same-sign top cross-section expected at the LHC 
(see e.g.~ref.~\cite{Berger:2011sv}). 
More exotic scenarios, with a $t$-channel exchange of more gauge bosons from non-Abelian horizontal symmetries 
may give rise to a sizable $A^{t\bar t}_{FB}$, while being consistent with the bounds on the same-sign top
cross section~\cite{Jung:2009jz,Kamenik:2011wt}. However, in such models there is no longer a clear correlation between
the non-standard contributions to $A^{t\bar t}_{FB}$ and $\Delta a_{CP}$. 

\subsection{New-physics scenarios with scalar-mediated FCNC}
\label{sect:Hmediated}

We finally analyze a new-physics framework with effective FCNC couplings to SM quarks of 
a scalar particle, which can be either the SM Higgs or some new scalar state.
In analogy to eq.~(\ref{eq:eff_lagr_Z}) we introduce the following 
effective Lagrangian
\be
\mathcal{L}^{h-{\rm FCNC}}_{\rm eff} = -
{\bar q}_i
\left[ (g^{h}_{L})_{ij} \, P_L + (g^{h}_{R})_{ij} \, P_R \right] q_j ~ h
+ \text{ h.c.}\,,
\label{eq:eff_lagr_h}
\ee
where $h$ is the scalar state. For simplicity, we assume that $h$ is a mass eigenstate 
and a $SU(2)_L\times U(1)_Y$ singlet. The scalar field $h$ could be identified with the physical Higgs boson, for instance in models with non-renormalizable interactions between quarks and multiple powers of the Higgs field~\cite{multiggs}. In this case, $\mathcal{L}^{h-{\rm FCNC}}_{\rm eff}$ results from these non-renormalizable 
interactions, after spontaneous electroweak breaking and diagonalization of the quark mass terms. 
In general, assuming $h$ to be a $SU(2)_L\times U(1)_Y$ singlet implies that  
$\mathcal{L}^{h-{\rm FCNC}}_{\rm eff}$
breaks explicitly the electroweak symmetry. It is then natural to normalize its
effective couplings as follows
\be
(g^{h}_{L})_{ij} = \frac{v}{M_{\rm NP}} (\lambda^h_L)_{ij}\, ,
\qquad
(g^{h}_{R})_{ij} = \frac{v}{M_{\rm NP}} (\lambda^h_R)_{ij}\, ,
\ee
where the $(\lambda^h_{L,R})_{ij}$ are dimensionless flavor off-diagonal terms. However, in models where $h$ is identified with the Higgs boson, we expect that $g^{h}_{L,R}\propto (v/M_{\rm NP})^n$, with $n$ an even integer. 

Also in this case the  chromomagnetic operator is generated at the one-loop level, with a  
leading contribution from $h$--top exchange diagrams. This leads to 
\be
C_{8} = \frac{\sqrt{2}}{4G_F} \frac{(g^{h}_{L})^*_{ut} (g^{h}_{R})_{tc}}{m_h^2} \frac{m_t}{m_c} f_8(x_{th})~, \qquad 
C_{7} = Q_u C_{8}~,
\ee
where $x_{th} = m^2_t/m^2_h$, ${\tilde C}_{7,8}$ are obtained via the replacement
$L \leftrightarrow R$, and 
\be
 f_8(x) =  \frac{x-3}{4(1-x)^2}-\frac{\log x}{2(1-x)^3}~.
\ee
Assuming $m_h=125$~GeV, the numerical 
expressions for  $\Delta a_{CP}$ is then 
\be
\left|
\Delta a^{h-{\rm FCNC}}_{CP} \right| \approx 0. 6 \%
\left| \frac{ \mathrm{Im}
\left[(g^{h}_{L})^*_{ut} (g^{h}_{R})_{tc}\right]  }{2 \times 10^{-4}} \right|
\approx 0.6\% \left| \frac{ \mathrm{Im}\left[(\lambda^{h}_{L})^*_{ut}(\lambda^{h}_{R})_{ct}\right] }{2\times 10^{-3}} \right|
\left(\frac{\rm 1~TeV}{M_{\rm NP}}\right)^2~.
\label{eq:acphNP}
\ee
Comments similar to those about eq.~(\ref{eq:acpZNP}) apply. However, in this case the different
$SU(2)_L\times U(1)_Y$ breaking structure implies a slower decoupling for large $M_{\rm NP}$.

Also the discussion about the bounds from $D^0-\bar{D}^0$  and $B_{s,d}$--$\bar B_{s,d}$
mixing proceeds in a similar way to the $Z$ FCNC case and will not be repeated here. 
The two main points can be summarized as follows: i)~$D^0-\bar{D}^0$ mixing constraints are satisfied once we forbid tree-level
contributions, and this is achieved assuming vanishing off-diagonal couplings between the first two generations;
ii)~$B_{s,d}$--$\bar B_{s,d}$ mixing constraints are satisfied assuming a mild hierarchy between
the effective couplings $(g^{h}_{L,R})^{ij}$ in the down and in the up sector, and also in this case 
the down-type couplings cannot be set to zero since they are radiatively induced  via Yukawa interactions.

As in all frameworks giving rise to an enhanced chromomagnetic operator,
the most severe constraints are posed by the (unavoidable)
contributions to the hadronic EDMs. In this case we have 
\be
d^{c}_{u} =-\frac{1}{8\pi^2}
\frac{\mathrm{Im}\left[(g^{h}_{L})^*_{ut} (g^{h}_{R})_{tu}\right]}{m_h^2} m_t f_8(x)~,
\qquad
\frac{d_{u}}{e}=Q_u d^{c}_{u}~,
\label{eq:edm_h}
\ee
and therefore
\be
|d_n| \approx 3 \times 10^{-26}~\left|
\frac{  \mathrm{Im}\left[(g^{h}_{L})^*_{ut} (g^{h}_{R})_{tu}\right]  }{2\times 10^{-7}} \right|
~{e\,\rm{cm}}\,,
\label{eq:dHNP}
\ee
for $m_h=125$~GeV.


With scalar-mediated FCNCs, the potentially most interesting signal are the rare top decays 
$t\to ch$ or $t\to uh$, if kinematically allowed. 
In particular, we find that
\be
\mathrm{Br}(t \to qh) \approx 0.4\times 10^{-2}
\left| \frac{(g^{h}_{R})^{tq}}{10^{-1}}\right|^2\,,
\label{eq:tqh}
\ee
which could be within the reach of the LHC (see fig.~\ref{fig:model_topFCNC}, where we assume $m_h=125$~GeV) 
although the observability of the signal depends on the specific decay modes of $h$.

A $t$-channel exchange of a relatively light scalar $h$ can contribute to
$A^{t\bar t}_{FB}$ and potentially decrease the tension between SM and data
(see e.g.~ref.~\cite{Blum:2011fa}). However, if  $h$ is a $SU(2)_L\times U(1)_Y$
singlet we face the same problem encountered with the $Z'$ from a $U(1)$
symmetry:  an excessive same-sign top cross-section expected at the LHC.
This fact can be understood quite easily: if the $u \to th$ coupling 
is allowed and $h$ is a self-conjugate field, then the $t$-channel exchange of $h$ 
leads to both $u \bar u \to t \bar t$ (relevant for $A^{t\bar t}_{FB}$ at the Tevatron)
and $u   u \to t   t$ (yielding a same-sign top cross section, particularly relevant at the LHC
given the large $uu$ parton component in $pp$ collisons). 
This problem can be avoided if we assume that $h$ is not an $SU(2)_L\times U(1)_Y$
singlet, as in ref.~\cite{Blum:2011fa}. However, 
in this case the one-loop contribution to the $\Delta C=1$ chromomagnetic 
operator is suppressed (additional electroweak symmetry-breaking terms are needed)
and, most important,  we loose a clear correlation between
the non-standard contributions to $A^{t\bar t}_{FB}$ and $\Delta a_{CP}$. 
A correlation is present with the contributions to $\Delta a_{CP}$ generated by 
$h$-mediated four-fermion interactions, as discussed in ref.~\cite{Hochberg:2011ru}.
However, according to the general analysis in ref.~\cite{Isidori:2011qw}, in this case 
the constraints from $\epsilon'/\epsilon$ do not allow to reach sizable values of $\Delta a_{CP}$.

\section{Conclusions}

It is not easy to assess whether new physics is necessary to explain the evidence for CP violation in charm, observed by LHCb through the difference in the time-integrated asymmetries in the decays $D^0 \to K^+K^-$ and $D^0\to \pi^+ \pi^-$~\cite{lhcb}. The central value of the recent measurement  is well above all expectations based on treating the charm as a heavy quark from the point of view of QCD interactions. To explain this observation within the SM, $D$ mesons should behave more like kaons rather than $B$ mesons, with non-perturbative enhancements yielding hadronic matrix elements 5-10 times larger than the corresponding estimates based on perturbative QCD. The lack of rigorous tools to evaluate these matrix elements does not allow us to exclude this possibility, although it goes against several other observations in the $D$ system, where the charm seems to behave as a heavy quark. It is therefore interesting to address the question of which extensions of the SM could account for a sizable fraction of the observed asymmetry.

If we assume that the observed CP violation in charm is caused by some new physics at the weak scale, we can deduce some clear and important lessons about the flavor structure of the new interactions. At the effective-theory level, consistency with other measurements (especially $D^0-{\bar D}^0$ mixing and $\epsilon^\prime /\epsilon$) strongly hints towards a $\Delta C=1$ chromomagnetic operator with large imaginary coefficient. As explained in sect.~1, selection rules offer a simple rationale for this choice.  Four-fermion $\Delta C=1$ effective interactions are disfavored, since they typically predict unacceptably large effects in other physical observables. Instead, the chromomagnetic operator, while giving a direct effect in singly-Cabibbo-suppressed decays, generates contributions to $D^0-{\bar D}^0$ mixing and $\epsilon^\prime /\epsilon$ that are always suppressed by at least the square of the charm Yukawa couplings, thus naturally explaining why they have remained undetected.

Some clear indications on the flavor pattern emerge also when we go beyond the effective theory and resolve the physics at the weak scale. In the case of a supersymmetric theory, the necessary ingredient is that the primary source of flavor violation comes from large left-right squark mixings. We have identified two flavor structures that can achieve this situation, successfully explaining CP violation in charm without any conflict with other experimental data. The first structure is what we call {\it disoriented $A$-terms}. This assumes flavor universality in squark masses and trilinear terms which are proportional to the corresponding Yukawa matrix, up to flavor-dependent coefficients of order unity. The large $\Delta C =1$ left-right transition is achieved by direct mixing between up and charm squarks of different chirality, with a coefficient proportional to $\lambda m_c$. The second structure exploits the large chiral transition of the top quark and it is naturally realized in models with {\it split families}, where the first two generations of squarks are much heavier than the third one. Taking advantage of the $m_t/m_c$ enhancement, small 1--3 and 2--3 mixings in the left-left and right-right sectors are sufficient to induce a chromomagnetic operator with the required coefficient, without causing problems in $\Delta C =2$ or $\Delta S =1$ processes. 

Both scenarios can be motivated by underlying model building. Split families have been considered for many years~\cite{io,altri}, but they have recently enjoyed special popularity after the LHC has cornered the minimal version of the supersymmetric model. Disoriented $A$-terms are more of a novelty, but there are good theoretical reasons for their existence and their implementations deserve more attention from model builders.

Both scenarios share some similar phenomenological aspects. Since the large left-right transition is one
of their fundamental ingredients, a near-maximal stop mixing is an expected consequence. This implies a relatively heavy Higgs boson. Indeed, our numerical analysis shows that a Higgs boson with mass around 125~GeV (as suggested by recent LHC findings) is fairly consistent with the measurement of $\Delta a_{CP}$. In both scenarios the dominant constraints are posed by the neutron and nuclear EDM, which are expected to be close to their experimental bounds. This result is fairly robust because the Feynman diagram contributing to quark EDMs has essentially the same structure as the one contributing to the chromomagnetic operator.
The EDM bounds require some tuning of the CP-violating phases of the models: a mild tuning in the case
of disoriented $A$-terms, thanks to the natural $m_{u,d}/m_c$ ratio between EDM and $\Delta a_{CP}$ contributions in that framework, and a more severe tuning in the split-family case, where similar flavor-mixing terms appear in both observables.
However, the specific tuning needed in the split-family case can be realized in models of alignment, where the ratio of the mixing terms in the up-type right-handed sector is related to $m_u/m_c$.
The two scenarios have distinct and particular predictions for other flavor-violating processes in $K$ and $B$ physics; in some cases, they could solve the tension between $\sin (2\beta)_{\rm tree}$ and $S_{\psi K_S}^{\rm exp}$. In the case of split families, there could be interesting effects in flavor-violating processes involving the top quark or the stops.

Our results have a more general validity than simply supersymmetry. In any new-physics model, the crucial ingredient to induce a chromomagnetic operator of the required size is the existence of a primary source of flavor violation in left-right transitions either direct (as in disoriented $A$-terms) or through the top Yukawa coupling (as in split families). In sect.~5 we have  illustrated this fact considering  non-supersymmetric models that meet these conditions. In particular, we have considered models with FCNC effective couplings of neutral gauge bosons ($Z$ or $Z'$) and scalar particles.
With suitable choices of the flavor-violating couplings, these modes can become very similar to the split-family supersymmetric scenario, as far as $\Delta a_{CP}$ and other low-energy observables are concerned. Then, not surprisingly, within these models it is possible to generate a large $\Delta a_{CP}$
provided the CP-violating phases are tuned to satisfy the tight neutron and mercury EDM bounds. Contrary
to the supersymmetric case, in these frameworks FCNC decays of the top occur at the tree level and may possibly be within the reach of the LHC. On the other hand, it is fair to say that none of the
non-supersymmetric frameworks we have considered satisfy the $\Delta F=2$ bounds as naturally as the
two supersymmetric scenarios.

On one side, our study has identified special theoretical structures that can explain the observed CP violation in charm, thus stimulating flavor model building. On the other side, we have discussed several physical observables that can provide new hints for determining whether the LHCb result can or cannot be explained by the SM.

\subsection*{Acknowledgments}
We thank Y. Grossman, G. Perez, L. Randall, and R. Rattazzi for useful discussions.
This work was supported by the EU ERC Advanced Grant FLAVOUR (267104),
and by MIUR under contract 2008XM9HLM.
G.I. acknowledges the support of the Technische Universit\"at M\"unchen -- Institute for Advanced
Study, funded by the German Excellence Initiative.

\end{document}